\documentclass{article}

\usepackage{amssymb,amsfonts,amsmath}
\usepackage{graphicx}
\usepackage{subfig}
\usepackage{bm}
\usepackage{cite}
\usepackage{xr}
\usepackage{authblk}

\usepackage[usenames,dvipsnames]{xcolor}

\begin{document}

\title{Genome-wide modelling of transcription kinetics reveals patterns of RNA production delays}

\author[1]{Antti Honkela}
\affil[1]{Helsinki Institute for Information
    Technology HIIT, Department of Computer Science, University of
    Helsinki, Finland}
\author[2,3]{Jaakko Peltonen}
\affil[2]{Helsinki Institute for Information
    Technology HIIT, Department of Computer Science,
    Aalto University, Espoo, Finland}
\affil[3]{School of Information Sciences, University of Tampere, Finland}
\author[2]{Hande Topa}
\author[4]{Iryna Charapitsa}
\affil[4]{Institute for Molecular Biology, Mainz, Germany}
\author[5]{Filomena Matarese}
\affil[5]{Nijmegen Centre for Molecular
    Life Sciences, Radboud University Nijmegen, NL}
\author[6]{Korbinian Grote}
\affil[6]{Genomatix Software GmbH, Muenchen, Germany}
\author[5]{Hendrik G. Stunnenberg}
\author[4]{George Reid}
\author[7]{Neil D. Lawrence}
\affil[7]{Department of Computer Science,
    University of Sheffield, UK}
\author[8]{Magnus Rattray}
\affil[8]{Faculty of Life Sciences,
    University of Manchester, UK}
\maketitle

\begin{abstract}
  Genes with similar transcriptional activation kinetics can display
  very different temporal mRNA profiles due to differences in
  transcription time, degradation rate and RNA processing
  kinetics. Recent studies have shown that a splicing-associated
  RNA production delay can be significant. To investigate this issue
  more generally it is useful to develop methods applicable
  to genome-wide data sets. We introduce a joint model of
  transcriptional activation and mRNA accumulation which can be used
  for inference of transcription rate, RNA production delay and
  degradation rate given data from high-throughput sequencing time
  course experiments.  We combine a mechanistic differential equation
  model with a non-parametric statistical modelling
  approach allowing us to capture a broad range of activation
  kinetics, and use Bayesian parameter estimation to quantify the
  uncertainty in 
  the estimates of the kinetic parameters.
  We apply the model to data from estrogen receptor (ER-$\alpha$)
  activation in the MCF-7 breast cancer cell line.
  We use
  RNA polymerase II (pol-II) ChIP-Seq time course data to characterise
  transcriptional activation and mRNA-Seq time course data to quantify mature
  transcripts. We find that
  11\% of genes with a good signal in the data display a delay of more
  than 20 minutes between completing transcription and mature mRNA
  production. The genes displaying these long delays are significantly
  more likely to be short.  We also find
  a statistical association between high delay and late
  intron retention in pre-mRNA data, indicating significant
  splicing-associated production delays in many genes.
\end{abstract}

\section{Introduction}

Induction of transcription through extracellular signalling
can yield rapid changes in gene expression for many genes.
Establishing the timing of events during this process is important for
understanding the rate-limiting mechanisms regulating the response and
vital for inferring causality of regulatory events.  Several processes
influence the patterns of mRNA abundance observed in the cell,
including the kinetics of transcriptional initiation, elongation,
splicing and mRNA degradation.  It was recently demonstrated that
significant delays due to the kinetics of splicing can be an important
factor in a focussed study of genes induced by Tumor Necrosis Factor
(TNF-$\alpha$)~\cite{Hao2013}. Delayed transcription can play an
important functional role in the cell; for example, inducing
oscillations within negative feedback loops~\cite{Monk2003} or
facilitating "just-in-time" transcriptional programmes with optimal
efficiency~\cite{Zaslaver2004}. It is therefore important to identify
such delays and to better understand how they are regulated. In this
contribution we combine RNA polymerase (pol-II) ChIP-Seq data with
RNA-Seq data to study transcription kinetics of estrogen receptor
signalling in breast cancer cells.  Using an unbiased genome-wide
modelling approach we find evidence for large delays in mRNA
production in 11\% of the genes with a quantifiable signal in our
data. A statistical analysis of genes exhibiting large delays
indicates that splicing kinetics is a significant factor and can be
the rate-limiting step for gene induction.

A high-throughput sequencing approach is attractive as it
gives broad coverage and thus allows us to uncover the typical
properties of the system. However, high-throughput data are associated
with significant sources of noise and the temporal resolution of our
data is necessarily reduced compared to previous studies using more
focussed PCR-based assays~\cite{Hao2013, Zeisel2011}. We have
therefore developed a statistically efficient model-based approach for
estimating the kinetic parameters of interest. We use Bayesian
estimation to provide a principled assessment of the uncertainty in
our inferred model parameters.
Our model can be applied to all genes with sufficiently strong
signal in both the mRNA and pol-II data with only mild restrictions on
the shape of the transcriptional activation profile (1814 genes here).

A number of other works studying transcription and splicing dynamics
(e.g.~\cite{Khodor2012,Pandya-Jones2013,Hao2013}) forgo detailed
dynamical modelling, which limits their ability to properly account
for varying mRNA half-lives.
Our statistical model incorporates a linear ordinary differential
equation of transcription dynamics, including mRNA degradation.
Similar linear differential equation models have been proposed as models of
mRNA dynamics previously~\cite{Rabani2011,Zeisel2011,Martelot2012},
but assuming a specific parametric form for the transcriptional activity.
In contrast, we apply a non-parametric Gaussian process framework
that can accommodate a quite general shape of transcriptional activity.  As
demonstrated previously~\cite{Lawrence2007,Gao2008,Honkela2010}, the
linearity of the differential equation allows efficient exact Bayesian
inference of the transcriptional activity function. Before presenting our 
results we outline our modelling approach.

\section{Model-based inference of transcriptional delays}

\begin{figure*}[t]
  \centering
  \includegraphics[width=\textwidth]{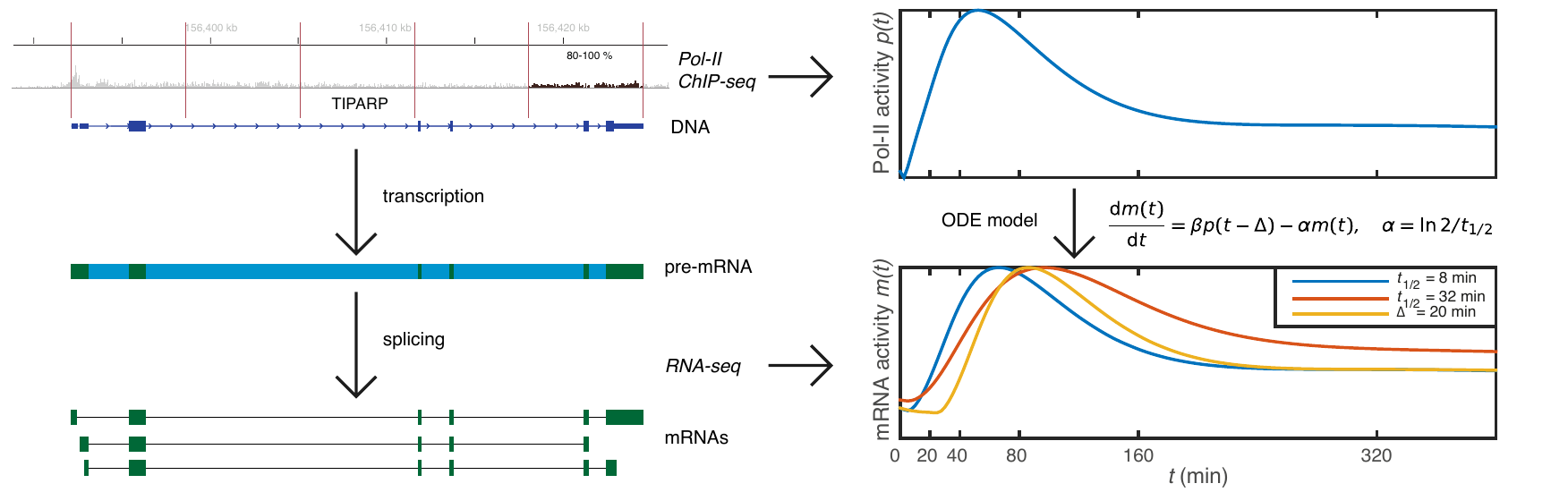}
  \caption{A cartoon illustrating the underlying biology and data
    gathering at a single time point (left) and time series modelling
    (right). The data come from pol-II ChIP-seq, summarised over the
    last 20\% of the gene body, and RNA-seq computationally split to
    pre-mRNA and different mRNA transcript expression levels. The
    modelling on the right shows the effect of changing mRNA
    half-life ($t_{1/2}$) or RNA production delay ($\Delta$) on the
    model response: both induce a delay on the mRNA peak
    relative to the pol-II peak, but the profiles have otherwise
    distinct shapes.}
  \label{fig:cartoon}
\end{figure*}

Our modelling approach is summarised in Fig.~\ref{fig:cartoon}.
We model the dynamics of transcription using a linear differential equation,
\begin{equation}
\frac{\mathrm{d}m(t)}{\mathrm{d}t} = \beta p(t-\Delta) - \alpha m(t) \ ,
\label{eqn:model}
\end{equation}
where $m(t)$ is the mature mRNA abundance and $p(t)$ is the
transcription rate at the 3' end of the gene at time $t$ which
is scaled by a parameter
$\beta$ since we do not know the scale of our $p(t)$ estimates. The parameter
$\Delta$ captures the delay between transcription completion and
mature mRNA production. We refer to this as the RNA production delay, defined as the time 
required for the polymerase to disengage from the 
pre-mRNA and be fully processed into a mature transcript. The parameter 
$\alpha$ is the mRNA degradation
rate which determines the mRNA half-life ($t_{\tiny 1/2} = \ln
2/\alpha$). We infer all model parameters ($\alpha$, $\beta$,
$\Delta$, the noise variance and parameters of the Gaussian process
covariance function discussed below) using a Markov chain Monte Carlo (MCMC)
procedure. The posterior distribution of the model
parameters quantifies our uncertainty and we use percentiles of the
posterior distribution when reporting credible regions around the mean
or median values.

\begin{figure*}[t]
  \centering
  \includegraphics[width=\textwidth]{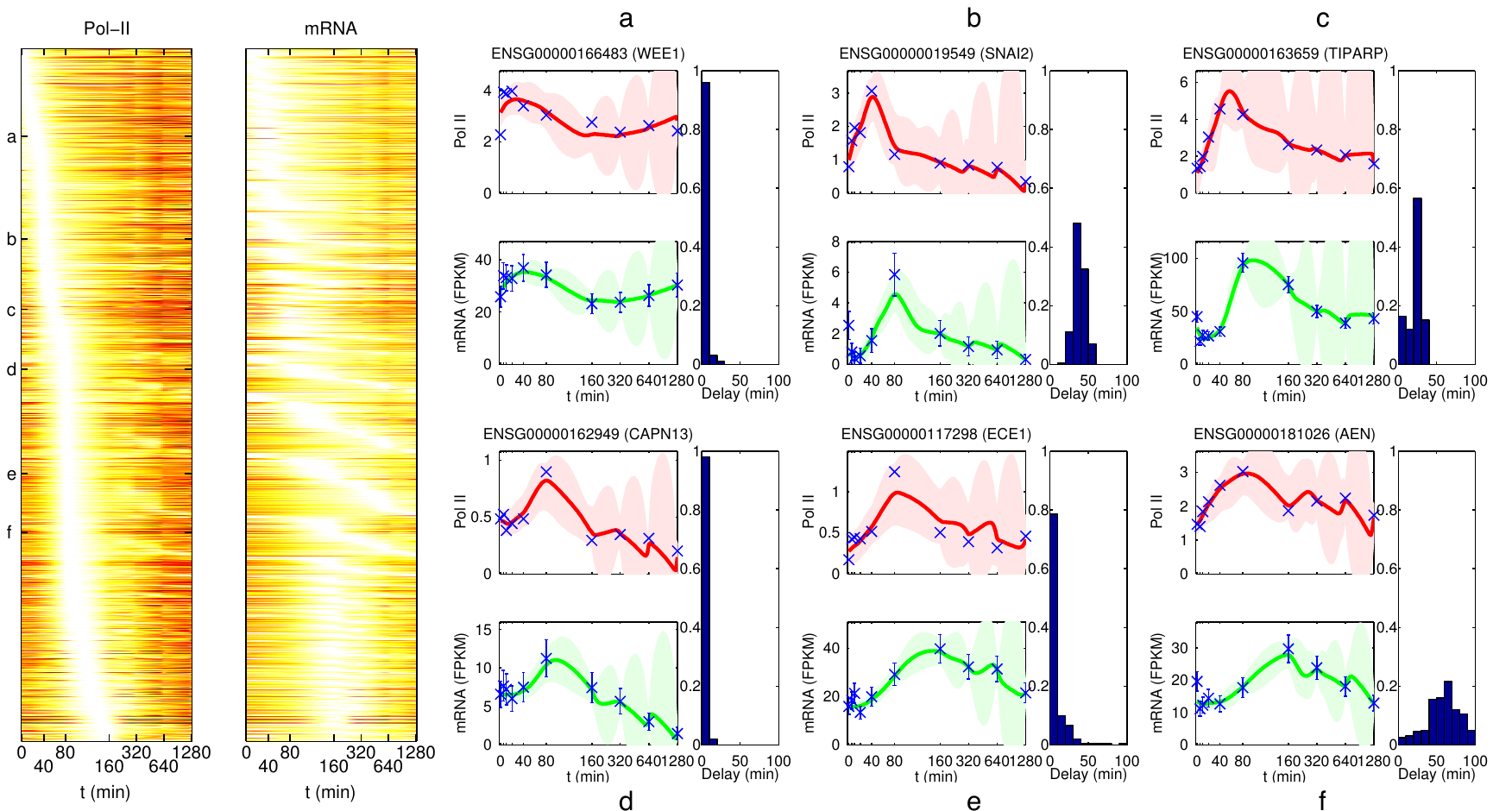}
  \caption{Left: Heat map of inferred pol-II and mRNA activity
    profiles after MCF7 cells are stimulated with estradiol. Genes with sufficient signal 
for modelling are sorted by the time of peak pol-II 
activity in the fitted model.
    Right: Examples of fitted model for six genes. For each gene, we show the fit
    using the pol-II ChIP-Seq data (collected from the final 20\% of
    the transcribed region) representing the transcriptional activity
    $p(t)$ (see Eqn.~\eqref{eqn:model}), and using the RNA-seq
    data to represent gene expression $m(t)$.
    Solid red/green lines show the mean model estimates for the
    pol-II/mRNA profiles respectively with associated credible regions.
    In each case we show the posterior distribution for the inferred
    delay parameter $\Delta$ to the right of the temporal profiles.
    Note that the final measurement times are very far apart (the $x$-axis is
    compressed to aid visualisation) leading to high uncertainty in
    the model fit at late times. However, this does not significantly
    affect the inference of delays for early induced genes.
    \label{fig:models1} }
\end{figure*}

We measure the transcriptional activity $p(t)$ using RNA polymerase (pol-II)
ChIP-Seq time course data collected close to the 3' end of the gene
(reads lying in the last 20\% of the transcribed region). Our main assumption is that 
pol-II abundance at the 3' end of the gene is proportional to 
the production rate of mature mRNA after a possible delay $\Delta$ due to 
disengaging from the polymerase and processing. The mRNA abundance is measured using RNA-Seq reads mapping
to annotated transcripts, taking all annotated transcripts into
account and resolving mapping ambiguities using a probabilistic
method~\cite{Glaus2012} (see Methods Section for details). As we limit
our analysis to pol-II data collected from the 3'-end of the
transcribed region, we do not expect a significant contribution to
$\Delta$ from transcriptional delays when fitting the model. Such transcriptional delays 
have recently been studied by modelling transcript elongation dynamics 
using pol-II ChIP-Seq time course 
data~\cite{waMaina2014} and nascent mRNA (GRO-Seq) data~\cite{Danko2013} in the
same system. Here we instead focus on production delays that can occur
after elongation is essentially complete.

Existing approaches to fitting models of this type have assumed a
parametric form for the activation function
$p(t)$~\cite{Rabani2011,Zeisel2011,Martelot2012}. We avoid restricting
the function shape by using a non-parametric Bayesian procedure for
fitting $p(t)$. We model $p(t)$ as a function drawn from a Gaussian
process which is a distribution over functions. The general properties
of functions drawn from a Gaussian process prior are determined by a
{\em covariance function} which can be used to specify features such
as smoothness and stationarity. We choose a covariance function that
ensures $p(t)$ is a smooth function of time since our data are
averaged across a cell population. Our choice of covariance function
is non-stationary and has the property that the function has some
persistence and therefore tends to stay at the same level between
observations (see Supplementary Material for further details). The
advantage of using a non-parametric approach is that we only have to
estimate a small number of parameters defining the covariance function
(two in this case, defining the amplitude and time-scale of the
function). If we were to represent $p(t)$ as a parametrised function
we would have to estimate a larger number of parameters to describe
the function with sufficient flexibility. The Bayesian inference
procedure we use to associate each estimated parameter with a credible
region would be more challenging with the inclusion of these
additional parameters.

We have previously shown how to perform inference over
differential equations driven by functions modelled using Gaussian
processes~\cite{Lawrence2007,Gao2008,Honkela2010}. The main methodological novelty in
the current work is the inclusion of the delay term in
equation~\eqref{eqn:model} and the development of a Bayesian inference
scheme for this and other model parameters. In brief, we cast the
problem as Bayesian inference with a Gaussian process prior
distribution over $p(t)$ that can be integrated out to obtain the data
likelihood under the model in Eqn.~\eqref{eqn:model} assuming
Gaussian observation noise. This likelihood function and its gradient are used for inference
with a Hamiltonian MCMC algorithm~\cite{Duane1987} to obtain a
posterior distribution over all model parameters and the full pol-II
and mRNA functions $p(t)$ and $m(t)$.

\section{Results}

We model the transcriptional response of MCF-7 breast cancer
cells after stimulation by estradiol to activate estrogen receptor (ER--$\alpha$) signalling. Fig.~\ref{fig:models1} shows 
the inferred pol-II and mRNA profiles for all genes with sufficient signal for modelling, along
with some specific examples of fitted models and estimated delay parameters.  Before discussing these results further below, we describe the application of our method to realistic simulated data to assess the reliability of our approach for parameter estimation under a 
range of conditions. 

\subsection{Simulated data}

\begin{figure}[t]
  \centering
  \includegraphics{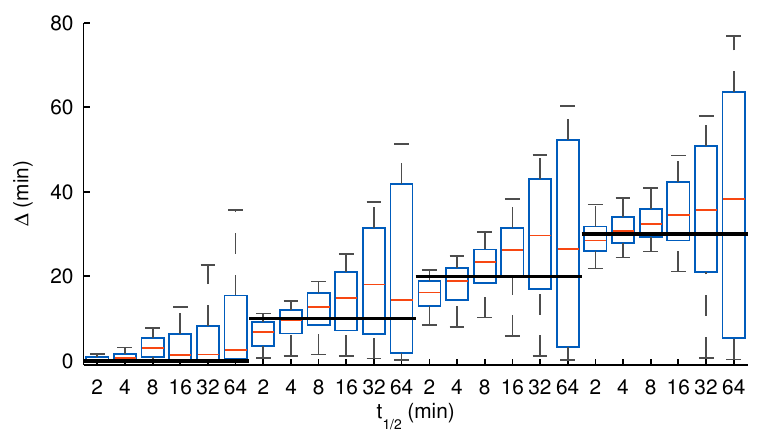}
  \caption{Boxplots of parameter posterior distributions illustrating
    parameter estimation performance on synthetic data for the delay
    parameter $\Delta$.  The strong
    black lines indicate the ground truth used in data generation.
    The box extends from 25th to 75th percentile of the posterior distribution while the whiskers
    extend from 9th to 91st percentile.
    The results show that delay estimates are accurate and reliable,
    with the true value always in the high posterior density region.
\label{fig:synthetic}
}
\end{figure}

We applied our method to data simulated from the model in
Eqn.~\eqref{eqn:model} using a $p(t)$ profile inferred using
pol-II data from the TIPARP gene (gene c in Fig.~\ref{fig:models1};
see Supplementary Material for further details over the simulated data).
We simulated data using different values of $\alpha$ and $\Delta$
to test whether we can accurately
infer the delay parameter $\Delta$. Fig.~\ref{fig:synthetic}
shows the credible regions of $\Delta$ for different ground truth
levels (horizontal lines) and for different mRNA degradation rates
(half-lives given on the $x$-axis). The results show that $\Delta$ can
be confidently inferred with the ground truth always lying within the
central part of the credible region.  The maximum error in posterior
median estimates is less than 10 min and when positive, the true value
is always above the 25th percentile of the posterior.
We observed that as the mRNA
half-life increases, our confidence in the delay estimates is
reduced. This is because the mRNA integrates the transcriptional
activity over time proportional to the half-life leading to a
more challenging inference
problem. We also note that inference of the degradation parameter
$\alpha$ is typically more difficult than inference of the delay
parameter $\Delta$ (see Fig.~\ref{fig:synthetic_halflives}).
However, a large uncertainty in the inferred
degradation rate does not appear to adversely affect the inference of
the delay parameters which are the main focus here.
More time-points, or a different spacing of time points,
would be needed to accurately infer the degradation rates.
Additional results of delay estimation in a scenario where the
simulated half-life changes during the time course are presented in
Fig.~\ref{fig:synthetic_changing}.  These results demonstrate that
the obtained delay estimates are reliable even in this scenario.

\subsection{Estrogen receptor signalling}

We applied our method to RNA-Seq and pol-II ChIP-Seq measurements from
MCF-7 cells stimulated with estradiol to activate ER-$\alpha$ signalling
(see Methods section). The
measurements were taken from cells extracted from the same population
to ensure that time points are directly comparable across
technologies. Example fits of our model are shown in
Fig.~\ref{fig:models1}. The examples show a number of different types
of behaviour ranging from early induced (a-c) to late induced (d-f),
and from very short delay (a, d, e) to longer delays (b, c, f).
Example (e), ECE1, is illuminating because visual inspection of the
profiles suggests a possible delay, but a more likely explanation
according to our model is a longer mRNA half-life and the posterior
probability of a long delay is quite low. Indeed, it is well known
that differences in stability can lead to delayed mRNA expression
\cite{Hao2009} and
therefore delays in mRNA expression peak relative to pol-II peak time 
are not sufficient to indicate a production delay. 
Changes in splicing can be another potential confounder, but our
transcript-based analysis of RNA-seq data can account for that.  An
example of how more naive RNA-seq analysis could fail here is
presented in Fig.~\ref{fig:osgin1_example}.

\begin{figure}[t]
  \centering
  \includegraphics{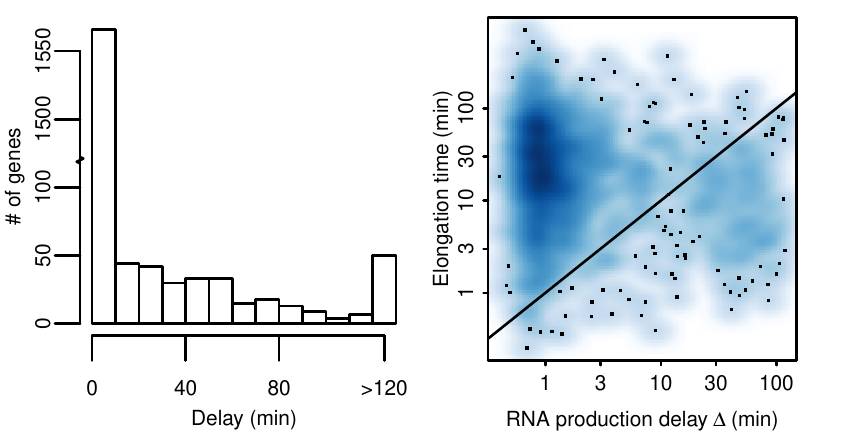}
  \caption{Left: A histogram of delay posterior medians from 1864 genes found
    to fit the model well.  Estimated delays
    larger than 120 min are
    considered unreliable and are grouped together.  These 50 genes were
    excluded from further analysis, leaving 1814 genes for the main
    analysis.
    Right: Estimated gene transcriptional delay for the longest
    transcript plotted against the estimated posterior median
    RNA production delay.  The transcriptional delay is estimated
    assuming each gene follows the median transcriptional velocity
    measured in Ref.~\cite{Danko2013}.  The solid line corresponds to
    equal delays.
  \label{fig:delay_analysis}}
\end{figure}

The parameter estimates of the models reveal a sizeable set of genes
with strong evidence of long delays between the end of transcription and 
production of mature mRNA.  We were able to obtain good model fits for
1864 genes.  We excluded 50 genes with posterior median delay >120
min, as these are unreliable due to sparse sampling late in
the time course, which is apparent from broad delay posterior
distributions.
Out of the remaining 1814 genes with reliable estimates,
204 (11\%) had a posterior median delay larger than 20 min between
pol-II
activity and mRNA production while 98 genes had the 25th percentile of delay posterior larger than 20 min,
indicating confident high delay estimates.  A histogram of 
median delays is shown in Fig.~\ref{fig:delay_analysis} (left).
The 120 min long delay cut-off was selected by visual observation of
model fits which were generally reasonable for shorter delays.
Note that late time points
in our data set are highly separated due to the exponential time
spacing used and thus the model displays high levels of
uncertainty between these points (see
Fig.~\ref{fig:models1}). Therefore genes displaying confident delay
estimates are typically early-induced such that time points are
sufficiently close for a confident inference of delay time. Our
Bayesian framework makes it straightforward to establish the
confidence of our parameter estimates.

\subsection{Genomic features associated with long-delay genes} 

\begin{figure}[t]
  \centering
  \includegraphics{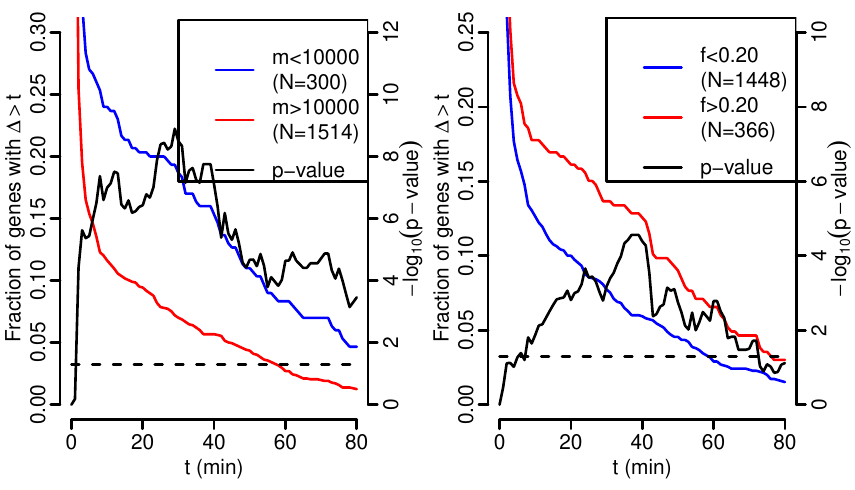}
  \caption{Tail probabilities for delays. Left: genes whose longest pre-mRNA transcript is short
    ($m$ is the length from transcription start to end). Right: genes with relatively long last introns ($f$ is the
    ratio of the length of the last intron of the longest annotated
    transcript of
    the gene divided by the length of that transcript pre-mRNA). The fraction of genes with long delays $\Delta$ is shown by
    the red and blue lines (left-hand vertical axis).
    In both subplots, the black curve denotes the $p$-values of
    Fisher's exact test for equality of fractions depicted by the
    red and blue curves
    conducted separately at each point (right-hand vertical
    axis) with the dashed line denoting $p<0.05$ significance
    threshold. Similar plots for other values of $m$ and $f$ as well
    as different gene filter setups are given in
    Figs.~\ref{fig:delay_tails1}--\ref{fig:delay_tails2}.
  \label{fig:delay_tails}}
\end{figure}

Motivated by previous studies~\cite{Khodor2012, Pandya-Jones2013, Bentley2014} we investigated 
statistical association
between the observed RNA production delay and genomic features related to
splicing. We found that genes with a short pre-mRNA
(Fig.~\ref{fig:delay_tails}, left panel) are more likely to have
long delays. We also find that genes where the ratio of the last
intron's length in the longest annotated transcript
over the total length of the transcript is large
(Fig.~\ref{fig:delay_tails}, right panel) are also more likely to have
long delays, but this effect appears to be weaker. These two
genomic features, short pre-mRNA and relatively long last introns, are
positively correlated, making it more difficult to separate their effects.
To do so, Fig.~\ref{fig:delay_tails_lenfilter} shows versions of
the right panel of Fig.~\ref{fig:delay_tails} but only including genes
with pre-mRNAs longer than 10 kb or 30 kb.  The number of genes with long last
introns in these sets is smaller and the resulting $p$-values are thus
less extreme, but the general shape of the curves is the same.
We did not find a significant relationship with the
absolute length of the last intron. This may be because the two
observed effects would tend to cancel out in such cases.
We also checked if exon skipping is associated with long delays as
previously reported~\cite{Pandya-Jones2013}.  The corresponding
results in Fig.~\ref{fig:delay_tails_exonskip} show no significant difference in
estimated delays in genes with and without annotated exon skipping.

\subsection{Analysis of the intronic read and pol-II distribution}

\begin{figure}[t]
  \centering
  \includegraphics{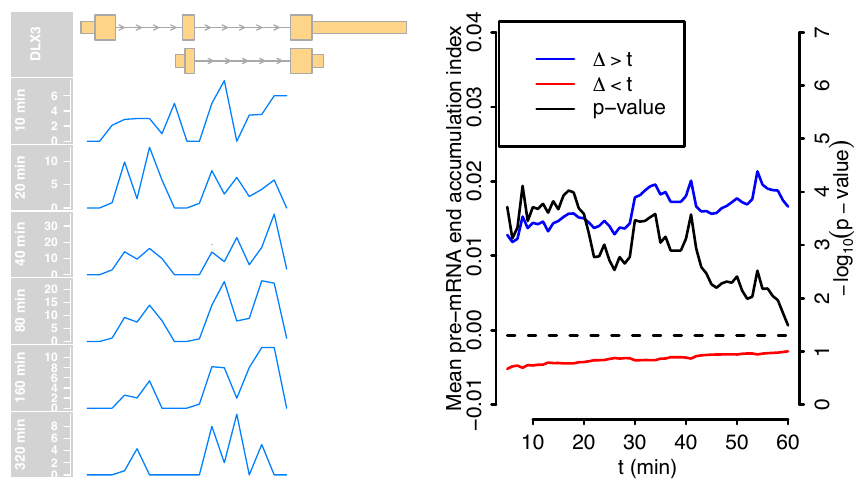}
  \caption{Left: We show the density of RNA-Seq reads uniquely mapping
    to the introns in the DLX3 gene, summarised in 200 bp bins. The
    gene region is defined from first annotated transcription start
    until the end of last intronic read. The ratio of the number of
    intronic reads after and before the midpoint of the gene region is
    used to quantify the 3' retention of introns. The pre-mRNA end
    accumulation index is the difference between averages of this
    ratio computed over late times (80-320 min) and early times
    (10-40 min). Right: Differences in the mean pre-mRNA
    accumulation index (left-hand vertical axis) in long delay genes (blue) and short delay
    genes (red) as a function of the cut-off used to distinguish the
    two groups (horizontal axis). Positive values indicate an increase in 3'
    intron reads over time. The black line shows the $p$-values of
    Wilcoxon's rank sum test between the two groups at each cut-off
    (right-hand vertical axis).
    \label{fig:premrna_accumulation}}
\end{figure}

We investigated whether there was evidence of differences in the
pattern of splicing completion for long-delay genes. To quantify 
this effect, we developed a pre-mRNA end accumulation index: 
the ratio of intronic reads in
the last 50\% of the pre-mRNA to the intronic reads in the first 50\%
at late (80-320 min) and early (10-40 min)
times. Fig.~\ref{fig:premrna_accumulation} shows that genes with a
long estimated delay display an increase in late intron retention at
the later times. There is a statistically significant difference
in the medians of index values for short and long delay genes
($p < 0.01$, Wilcoxon's rank-sum test $p$-values for different
short/long delay splits are shown in Fig.~\ref{fig:premrna_accumulation}).
The example on the left of Fig.~\ref{fig:premrna_accumulation},
DLX3, is a relatively short gene of about 5 kb and thus
differences over time cannot be explained by the time required for
transcription to complete. The corresponding analysis for pol-II
ChIP-seq reads as well as GRO-seq reads is in
Fig.~\ref{fig:pol2_accumulation}. It shows a clear delay-associated
accumulation to the last 5\% nearest to the 3' end, while for
pol-II in the last 50\% the accumulation is universal.
These results suggest our short delay genes tend to be
efficiently spliced while long delay genes are more likely to exhibit delayed
splicing towards the 3' end. There is also evidence of some
accumulation of pol-II near the 3' end although the effect appears relatively
weak. We note that Grosso \emph{et al.}~\cite{Grosso2012} identified genes
with elevated pol-II at the 3'-end which were found to be
predominantly short, consistent with our set of delayed genes, and
with nucleosome occupancy
consistent with pausing at the 3' end. 

\subsection{Relative importance of production and elongation delays}

To better understand what are the rate-limiting steps in transcription dynamics,
we assessed the relative importance of the observed RNA production delays in comparison 
to transcriptional delays due to elongation time.  We estimated elongation times for each gene using
assumed transcriptional velocity corresponding to the 2.1 kb/min
median estimate from~\cite{Danko2013} combined with the length of the
longest annotated pre-mRNA transcript.  Others (e.g.~\cite{waMaina2014}) have
reported higher velocities, so this approach should provide reasonable
upper bounds on actual elongation time for most genes.  A
comparison of these delays with our posterior median delay estimates
is shown in Fig.~\ref{fig:delay_analysis} (right).  The figure shows the
majority of genes with short production delays and moderate
elongation time in the top-left corner of the figure, but
14.3\% (260/1814) of genes have a longer RNA production delay 
than elongation time.

\section{Discussion}

Through model-based coupled analysis of pol-II and mRNA time course data
we uncovered the processes shaping mRNA expression changes in response to
estrogen receptor signalling. We find that a large number of genes
exhibit significant production delays. We also find that delays 
are associated with short overall gene length, relatively long final intron length and 
increasing late-intron retention over time. Our results support a  
major role for splicing-associated delays in shaping the timing of gene expression in 
this system.  Our study complements the discovery of similarly large splicing-associated delays in 
a more focussed study of TNF-induced expression~\cite{Hao2013} indicating that
splicing delays are likely to be important determinants of expression dynamics across a range of
signalling pathways. 

It is known that splicing can strongly influence the
kinetics of transcription.  Khodor {\em et al.} carried out a comparative study of splicing 
efficiency in fly and mouse and found a positive 
correlation between absolute gene length and splicing efficiency  \cite{Khodor2012}. This suggests that efficient co-transcriptional splicing is facilitated by increased gene length and is consistent 
with our observation that delays are more common in shorter genes. In these genes it 
appears that the mature mRNA cannot be produced after transcription until splicing is completed; it is splicing rather than transcription that is the rate-limiting step for these genes. In the same
study it was also observed that introns close to the 3'-end of a gene are less efficiently spliced which is consistent
with our observation that the relative length of the final intron may impact on splicing delays.  
A further theoretical model supporting a link between long final
introns and splicing inefficiency was recently suggested in
Ref.~\cite{Catania2013}, but it is unclear if it can fully explain
the observed relationships.

Our model assumes a constant mRNA degradation rate which may be
unrealistic.  Given the difficulty of estimating even a single
constant degradation rate for simulated data where the true rate is
constant, it seems infeasible to infer time-varying rates with the current
data. On the other hand, estimated delays were
quite reliably inferred even when we simulated data with a time-varying degradation rate (Fig.~\ref{fig:synthetic_changing}), and hence the potentially incorrect degradation model should not affect the main results significantly.

It is important to differentiate the delays found here with transcriptional
delays required for pol-II elongation to complete. Elongation time can be a significant factor
in determining the timing of gene induction and elongation dynamics has been modelled using both pol-II ChIP-Seq~\cite{waMaina2014} and nascent RNA (GRO-Seq)~\cite{Danko2013} time course measurements in the system considered here. However, in 
this study we limited our attention to pol-II data
at the 3'--end of the gene, i.e.\ measuring polymerase density changes in the region where elongation is almost completed. 
Therefore, we
will not see transcription delays
in our data and the splicing-associated delays discussed above are not related to elongation time. Indeed, the splicing-associated
delays observed here are more likely to affect shorter genes where transcription completes
rapidly. These splicing-associated delays are much harder to predict from genomic features than transcriptional delays, which are mainly determined by gene length, although
we have shown an association with final intron length and gene length. In the future it would be
informative to model data from other systems to establish associations with system-specific variables (e.g.\ alternative splice-site usage) and thereby uncover context-specific mechanisms regulating the 
delays that we have observed here.

\subsection{Availability} 
Raw data are available at GEO (accession GSE62789).
A browser of all model fits and delay estimates
is available at \texttt{http://ahonkela.users.cs.helsinki.fi/pol2rna/}.
Code for reproducing the experiments is available at\\
\texttt{https://github.com/ahonkela/pol2rna}.

\section{Methods}

\subsection{Data acquisition and mapping}

MCF-7 breast cancer cells were stimulated with estradiol (E2) after 
being placed in estradiol free media for three days,
similarly as previously described~\cite{waMaina2014}.  We measured pol-II
occupancy and mRNA concentration from the same cell population collected 
at 10 time points on a logarithmic scale: 0, 5, 10, 20, 40, 80, 160, 320, 640 and 1280 min 
after E2 stimulation. At each time point, the pol-II occupancy was measured genome-wide by
ChIP-seq and mRNA concentration using RNA-Seq. Raw reads from the ChIP-Seq data 
were mapped onto the human genome reference sequence (NCBI\_build37) using the Genomatix Mining Station
(software version 3.5.2; further details in Supplementary Material).
On average 84.0\% of the ChIP-Seq reads were mapped uniquely to the genome.
The RNA-seq reads were mapped using bowtie to a transcriptome
constructed from Ensembl version 68 annotation allowing at most 3
mismatches and ignoring reads with more than 100 alignments.  The
transcriptome was formed by combining the cDNA and ncRNA
transcriptomes with pre-mRNA sequences containing the full genomic
sequence from the beginning of the first annotated exon to the end of
the last annotated exon.  On average 84.7\% of the RNA-seq reads
were mapped.

\subsection{RNA-seq data processing}

mRNA concentration was estimated from RNA-seq read data using BitSeq \cite{Glaus2012}. BitSeq is a probabilistic method to infer transcript expression from 
RNA-seq data after mapping to an annotated transcriptome. We
estimated expression levels to all entries in the transcriptome,
including the pre-mRNA transcripts, and
used the sum of the mRNA transcript expressions in FPKM
units to estimate the mRNA expression level of a gene.
Different time points of the RNA-seq time series were normalised using
the method of \cite{Anders2010}.

\subsection{Pol-II ChIP-seq data processing}

The ChIP-seq data were processed into time series
summarising the pol-II occupancy at each time point for each human gene.
We considered the last 20\% of the gene body nearest to the 3'-end.
The gene body was defined from the start of the first exon to the end of
the last exon in Ensembl 68 annotation.  The data were subject to
background removal and normalisation of time points.
(Full details in the Supplementary Material.)

\subsection{Filtering of active genes}

We removed genes with no clear time-dependent activity by fitting
time-dependent Gaussian process models to the activity curves and only
keeping genes with Bayes factor at least 3 in favour of the
time-dependent model compared to a null model with no time dependence.
We also removed genes that had no pol-II observations at 2 or more
time points.  This left 4420 genes for which we fitted the models.

\subsection{Modelling and parameter estimation}

We model the relationship between pol-II occupancy and mRNA concentration
using the differential equation in Eqn.~\eqref{eqn:model} which relates the pol-II time 
series $p(t)$ and corresponding mRNA time series $m(t)$ for each gene. We model $p(t)$
in a nonparametric fashion by applying a Gaussian Process (GP) prior over the 
shapes of the functions.
We sightly modify the model in Eqn.~\eqref{eqn:model} by adding a constant $\beta_0$ to account for the limited depth of pol-II ChIP-Seq measurements,
\begin{equation}
\frac{\mathrm{d} m(t)}{\mathrm{d} t} = \beta_0 + \beta p(t-\Delta) -\alpha m(t)
\label{eq:differential_equation}
\end{equation}
This differential equation can be solved for $m(t)$ as a function of $p(t)$
in closed form.
The
pol-II concentration function $p(t)$ is represented as a sample from
a GP prior which can be integrated out to compute the data 
likelihood. The model 
can be seen as an extension of a previous model applied to
transcription factor target identification \cite{Honkela2010}.
Unlike Ref.~\cite{Honkela2010}, we
model $p(t)$ as a GP defined as an integral of a function
having a GP prior with RBF
covariance, which implies that $p(t)$ tends to remain constant
between observed data instead of reverting back to the mean.
Additionally we introduce the delay between pol-II
concentration and mRNA production as well as model the initial
mRNA concentration as an independent parameter.
In the
special case where 
$\Delta=0$, and $m_0=\beta_0/\alpha$,
Eqn.~\eqref{eq:mrna_integral_appendix} reduces to the previous model (Eqn.\ 4 in
\cite{Honkela2010}).
In order to fit the model to pol-II and mRNA
time course data sampled at discrete times, we assume
we observe $m(t)$ and $p(t)$ corrupted by zero-mean Gaussian noise
independently sampled for each time point. We assume
the pol-II noise variance is a constant $\sigma_{p}^2$ and infer
it as a parameter of the model. The mRNA noise variances
for each time point are sums of a shared constant $\sigma_{m}^2$ and
a fixed variance inferred by BitSeq by combining the technical quantification
uncertainty from BitSeq expression estimation with an estimate of
biological variance from the BitSeq differential expression model
(full details in Supplementary Material).

Given the differential equation parameters, GP inference yields a full
posterior distribution over the shape of the Pol-II and mRNA
functions $p(t)$ and $m(t)$. We infer the differential equation
parameters from the data using MCMC sampling which
allows us to assign a level of 
uncertainty to our parameter estimates. To infer a full
posterior over the differential equation parameters $\beta_0$,
$\beta$, $\alpha$, $\Delta$, $m_0$, $E[p_0]=\mu_p$,
the observation model parameters $\sigma_{p}^2$, $\sigma_{m}^2$,
and a magnitude parameter $C_p$ and width parameter $l$ of the GP prior, 
we set near-flat priors for them over reasonable value ranges, except for the
delay $\Delta$ whose prior is biased toward 0 (exact ranges and full
details are presented in Supplementary Material).
We combine these priors with the likelihood obtained from the GP model
after marginalising out $p(t)$ and $m(t)$, which can be performed
analytically.  We infer the
posterior over the parameters by Hamiltonian MCMC sampling.
This full MCMC approach utilises gradients of the distributions for
efficient sampling and  rigorously takes uncertainty over differential
equation parameters into account. Thus the final posterior accounts
for both the uncertainty about differential equation parameters, and
uncertainty over the underlying functions for each differential
equation.
We ran 4 parallel chains starting from different random initial states
for convergence checking using the potential scale reduction
factor of~\cite{Gelman1992}.  We obtained 500 samples from each of the
4 chains after discarding the first half of the samples as burn-in and
thinning by a factor of 10.
Posterior distributions over the functions $p(t)$ and $m(t)$ are
obtained by sampling 500 realisations of $p(t)$ and $m(t)$ for each
parameter sample from the exact Gaussian conditional posterior given
the parameters in the sample.  The resulting posteriors for $p(t)$ and
$m(t)$ are non-Gaussian, and are summarised by posterior mean and
posterior quantiles.
Full details of the MCMC procedure are in Supplementary Material.

\subsection{Filtering of results}

Genes satisfying the following conditions were kept for full
analysis. (Full implementation details of each step are in
Supplementary Material.)
\begin{enumerate}
\item $p(t)$ has the maximal peak in the densely sampled region between
  1 min and 160 min.
\item Estimated posterior median delay is less than 120 min.
\item $p(t)$ does not change too much before $t=0 \text{ min}$ to
  match the known start in steady state.
\end{enumerate}

\subsection{Analysis of the gene annotation features associated with the delays}

Ensembl version 68 annotations were used to derive features of all
genes.  For each annotated transcript, we computed the total pre-mRNA
length $m$ as the distance from the start of the first exon to the end
of the last exon, and the lengths of all the introns.  Transcripts
consisting only of a single exon (and hence no introns) were excluded
from further analysis.  For each gene, we identified the transcript
with the longest pre-mRNA and used that as the representative
transcript for that gene.  The last intron share $f$ was defined as
the length of the last intron of the longest transcript divided by
$m$.

\subsection{Pre-mRNA end accumulation index}

For this analysis, we only considered reads aligning uniquely to
pre-mRNA transcripts and not to any mRNA transcripts.  We counted the
overlap of reads with 200 bp bins starting from the beginning of the
first exon of each gene ending with the last non-empty bin.  We
compute the fraction $r_{e,i}$ of all reads in the latter half of bins
in each sample $i$, and define the index as the difference of the
means of $r_{e,i}$ over late time points (80-320 min) and over early
time points (10-40 min).

\section*{Acknowledgments}

The work was funded by the European ERASysBio+ initiative project
``Systems approach to gene regulation biology through nuclear
receptors'' (SYNERGY) by the BBSRC (BB/I004769/2 to JP, MR and
NDL), Academy of Finland (135311 to AH and HT) and by the
BMBF (grant award ERASysBio+ P\#134 to GR; grant no. 0315715B to KG).
MR, NDL and KG were further supported by EU FP7 project RADIANT
(grant no. 305626), and AH and JP by the Academy of Finland
(grant nos. 252845, 259440, 251170).

\clearpage

\appendix

\renewcommand\thefigure{S\arabic{figure}}
\setcounter{figure}{0}
\renewcommand\thetable{S\arabic{table}}
\setcounter{table}{0}
\renewcommand\theequation{S\arabic{equation}}
\setcounter{equation}{0}

\newcommand{\isqwidth}{u_p}
\newcommand{\polmag}{C_p}
\newcommand{\rnamag}{\beta^2}
\newcommand{\polwidth}{l}
\newcommand{\polnoisevar}{\sigma_p^2}
\newcommand{\rnanoisevar}{\sigma_m^2}
\newcommand{\alldata}{\mathcal{D}}

\section*{Supplementary material}

In the following we provide details of data acquisition,
processing of RNA-seq data and filtering of active genes, processing
of pol-II ChIP-seq data, differential equation modelling of the
connection between pol-II and mRNA, Gaussian process based inference
of underlying time series, and summarisation and filtering of
results. We then provide an explanation of how synthetic data were used
to study accuracy of parameter estimation for mRNA half life, a
measure of mRNA decay in the differential equation model between
pol-II and mRNA. Lastly, we provide additional figures about tail
probabilities of delays for alternative result filtering choices, an
additional  figure about long posterior mean delays
with and without annotated exon skipping, and differences in 
pre-mRNA accumulation in short and long delay genes.

\section{Data acquisition}

MCF-7 breast cancer cells were treated with estradiol (E2). The cells
were put in estradiol free media for three days. This is defined media
devoid of phenol red (which is estrogenic) containing 2\% charcoal
stripped foetal calf serum. The charcoal absorbs estradiol but not
other essential serum components, such as growth factors. This
resulted in basal levels of transcription from E2 dependent genes. The
cells were then incubated with E2 containing media, which resulted in
the stimulation of estrogen responsive genes. Measurements were taken 
at logarithmically spaced time points 0, 5,
10, 20, \dots, 1280 minutes after E2 stimulation.

At each time point, the pol-II occupancy was measured genome-wide by
ChIP-seq. Raw reads were mapped onto the human genome
reference sequence (NCBI\_build37) using the Genomatix Mining Station
(software version 3.5.2). The mapping software is an index-based
mapper using a shortest unique subword index generated from the
reference to identify possible read positions. A subsequent alignment
step is then used to get the highest-scoring match(es) according to
the parameters used. We used a minimum alignment quality threshold of
92\% for mapping, reads were not trimmed. On average 84\% percent of
reads could be mapped uniquely.

At each time point, the pre-mRNA and mature mRNA abundances were
measured for each human gene by RNA-seq. Total RNA was isolated and subjected 
to rRNA depletion with the Ribo-Zero Magnetic Gold Kit and processed further 
for strand-specific RNA-seq. The RNA-seq reads were mapped using Bowtie to 
a transcriptome
constructed from Ensembl version 68 annotation allowing at most 3
mismatches and ignoring reads with more than 100 alignments.  The
transcriptome was formed by combining the cDNA and ncRNA
transcriptomes with pre-mRNA sequences containing the full genomic
sequence from the beginning of the first annotated exon to the end of
the last annotated exon.  On average 84.7\% of the RNA-seq reads
were mapped.

All the ChIP-seq and RNA-seq data are available from the NCBI Gene
Expression Omnibus under accession number GSE62789.

\section{RNA-seq data processing}

RNA-seq data were analysed at each time point separately using
BitSeq~\cite{Glaus2012}.  The reads were first mapped to human
reference transcriptome (Ensembl v68) using Bowtie version 0.12.7~\cite{Langmead2009}.
In order to separate pre-mRNA activity as well, we augmented the
reference transcriptome with pre-mRNA transcripts for each gene that
consisted of the genomic sequence from the beginning of the first
exon to the end of the last exon of the gene.

BitSeq uses a probabilistic model to probabilistically assign
multimapping reads to transcript isoforms~\cite{Glaus2012}, in our
case also including the pre-mRNA transcripts.  We obtained gene
expression estimates by adding the corresponding mRNA transcript
expression levels.  In addition to the mean expression levels, BitSeq
provides variances of the transcript isoform expression levels.  We
further used the biological variance estimation procedure from BitSeq
differential expression analysis on the estimated gene expression
levels by treating the first three time points (0, 5, 10 min) as
biological replicates.  Genes with similar mean expression levels
(log-RPKM) were grouped together such that each group contained 500
genes except for the last group with 571 genes with the highest
expression.  Then, the biological variances were estimated for each
group of genes by using the Metropolis--Hastings algorithm used in
BitSeq stage 2~\cite{Glaus2012}.  Biological variances for the single
measurements were determined according to the gene expression levels
at each time point, where each gene was considered to belong to the
closest gene group according to its expression level.  The observation
noise variance for each observation was defined as the sum of the
technical (BitSeq stage 1) and biological (BitSeq stage 2) variances,
and transformed from log-expression to raw expression using
\begin{equation}
  \label{eq:variance_transformation}
  \sigma^2_{\text{raw}} = \sigma^2_{\text{log}} \exp(\mu_{\text{log}})^2.
\end{equation}

Different time points of the RNA-seq time series were normalised using
the method of \cite{Anders2010} as implemented in the edgeR
R/Bioconductor package~\cite{Robinson2010}.

Statistics of RNA-seq mapping and distribution of reads for pre-mRNA
and mRNA transcripts are presented in Tables~\ref{tab:alignmentstats}
and \ref{tab:bitseqstats} as well as Fig.~\ref{fig:rnastats}.

\section{Filtering of active genes}

We removed genes with no clear time-dependent activity by fitting
time-dependent Gaussian process models to the activity curves and only
keeping genes with Bayes factor at least 3 in favour of the
time-dependent model compared to a null model with no time dependence.
We also removed genes that had no pol-II observations at 2 or more
time points.  This left 4420 genes for which we fitted the models.

\section{Pol II ChIP-seq data processing}

The ChIP-seq data were processed into time series by
summarising the pol-II occupancy over time for each human gene
(Ensembl version 68 annotation was used for gene positions), by a
series of steps as follows.
\begin{enumerate}
\item Each gene was divided into 200 bp bins and levels of pol-II activity
were computed at each time point 
as the total weighted count of reads overlapping each bin, where each
read was weighted by how many basepairs in the read overlap the bin, as follows. Only
uniquely mapped reads were used. For any read that at least partially
overlaps a bin, the number of basepairs overlapping the bin was added
into the activity level of the bin. For any read spanning multiple
bins, the number of basepairs overlapping each bin were added into activity of
that bin. The Genomatix mapping software provides alignment scores
(values between 0 and 1; with our threshold only between 0.92 and 1)
for mapping reads to the genome; for any read
having alignment score less than 1, the number of overlapping
basepairs added to each bin was multiplied by the alignment score.

\item A noise removal was then done: a noise level was computed as the
  average activity level in 74 manually selected regions from
  Chromosome 1 that were visually determined to be inactive over the
  measurement time points, as follows.  
  The regions were divided into 200 bp bins, and
  total weighted counts of reads overlapping each bin (each read
  weighted by the number of basepairs overlapping the bin) were
  computed in the same way as for the genes in the previous step.  For
  each time point, the noise level was computed as the average
  activity level over all bins from all 74 regions.  The computed
  noise level was subtracted from the mean of each bin in each gene,
  thresholding the result at zero.  A list of the empty regions
  used 
  is included as a Supplementary Dataset S1.

\item As the number of ChIP-seq reads collected overall for pol-II varies
between time points, a robust normalisation was done. After the
previous noise removal step, for each gene $g$ at each time $t$ 
we compute the mean of the remaining activity (activity level after
noise removal) over bins of the gene, denoted as $r_{gt}$.
The activity levels are weighted counts of basepairs from reads
overlapping the gene; we select genes having sufficient activity,
that is, at least $5\cdot 200$ overlapping basepairs from reads over 
each 200 bp bin of the gene, on average over the bins. For each gene $g$ let 
$T_{g} = \{t'\in \{5,10,\ldots,1280\mathrm{\;min}\} | r_{gt'}>5\cdot 200\}$ 
denote those time points (except the first time point) where the gene has
sufficient activity.
For each time point we
compute a normalisation factor of~\cite{Anders2010}
$$
     C_t = \textrm{Median}_g \left\{ \frac{r_{gt}}{\textrm{GeomMean}_{t'} \{r_{gt'}\} } \right\} \;.
$$
where $\textrm{Median}_g\{\cdot \}$ denotes median over genes and\linebreak
$\textrm{GeomMean}_{t'}\{r_{gt'}\}=(\prod_{t'\in T_g} r_{gt'})^{1/|T_g|}$ is the 
geometric mean over the time points having sufficient activity for
gene $g$. The median is computed for time points after the first time
point; for the first time point $t=0\mathrm{\;min}$
we set $C_t=1$.  
The factor $C_t$ normalises all the gene activity levels (weighted read counts) at a time point downwards if
genes at that time point have unusually many reads, exceeding their
(geometric) mean activity level, and normalises upwards if gene
activity levels fall
under their mean activity level.

\item Lastly, time series summaries were computed for pol-II at each gene.
For each gene at each time point $t$, the mean activity level
(weighted read-count) of pol-II
over bins in the 20\% section of the gene nearest to transcription end
was computed, normalised by $C_t$. This measured pol-II level
represents transcriptional activity that had successfully passed
through the gene to the transcription end site; it is expected to 
correspond better with mRNA production rate than pol-II activity at the
transcription start of the gene, since pol-II near the 
transcription start site
can be in the active or
 inactive state and after activation may require a significant time for 
transcription to complete. 

\item For a small number of genes where the active mRNA transcripts covered only
part of the gene, we considered the area from the first active exon to
last active exon, and summarised the gene using the 20\% section
nearest to the end of the area. Active transcripts were defined here
as transcripts with a mean of more than 1.1 assigned counts in the
BitSeq posterior expression estimates. BitSeq uses a prior that
assigns 1 ``pseudo-count'' per transcript, so the active transcripts
were only required to have minimal posterior expression that was
distinguishable from the prior.  A list of active transcripts
is included as Supplementary Dataset S2.

\item Lastly, for mathematical convenience, for each pol-II time series we
subtracted from all time points the minimum value over the time points.
\end{enumerate}

\section{Differential equation based modelling}

We model the role of pol-II as a catalyst of the transcription of DNA
into mRNA as a differential equation for each gene; the differential
equation relates the pol-II time series $p(t)$ of the gene and the
corresponding mRNA time series $m(t)$.

Let us assume the momentary pol-II activity directly represents the
momentary rate of transcription, potentially with a delay, and that
the mRNA decays at a constant rate. We model this as a
linear differential equation
\begin{equation}
\frac{\mathrm{d} m(t)}{\mathrm{d} t} = \beta_0 + \beta p(t-\Delta) -\alpha m(t)
\label{eq:differential_equation_appendix}
\end{equation}
where $\Delta$ is a delay parameter between the pol-II activity and
the momentary transcription rate, $\beta_0$ is a parameter representing the
\emph{baseline transcription rate from unobservable pol-II background}
(baseline production level of
mRNA), $\beta$ is a parameter representing \emph{transcriptional efficiency}, that is, 
sensitivity of the transcription rate to activity of pol-II, and $\alpha$ is a
constant mRNA \emph{decay} rate parameter that is related to
mRNA half-life $t_{1/2}$ through $\alpha = \ln(2) / t_{1/2}$.

The momentary mRNA level $m(t)$ can be solved from the differential
equation 
to yield the following solution:
\begin{equation}
m(t)
=m_0 e^{\alpha(t_0-t)} + 
\frac{\beta_0}{\alpha}\left(1-e^{-\alpha t}\right)
+ \beta e^{-\alpha t}\int_{u=0\mathrm{\;min}}^t e^{\alpha u} p(u-\Delta) \mathrm{d}u 
\label{eq:mrna_integral_appendix}
\end{equation}
where pol-II activity is assumed to start at $t=0\mathrm{\;min}$
($p(t)=0 \text{ for } t<0\mathrm{\;min}$), $t_0 \gg 0\mathrm{\;min}$ is the
time of the first observation, and $m_0$ is an initial mRNA abundance
at $t_0$ 
which is inferred as a parameter of the model. 
No parametric assumptions are made about the shape of the pol-II time
series function $p(t)$, and the only assumption about the mRNA level $m(t)$
is that it arises through the differential equation.  

The linear differential equation \eqref{eq:differential_equation_appendix} and
its linear solution operator \eqref{eq:mrna_integral_appendix} are similar to
those used previously in~\cite{Lawrence2007,Gao2008,Honkela2010}
except for the added delay.  As in the previous works, the linearity
of the solution operator permits exact joint Gaussian process (GP)
modelling over $p(t)$ and $m(t)$.

\section{Gaussian process inference}

We model pol-II and mRNA time series, $p(t)$ and $m(t)$, in a nonparametric fashion 
which avoids the 
assumption of a specific parametric shape for the time series function; 
instead, we set a GP prior over the time series functions.

For each gene, GP inference of the posterior distribution over the underlying 
pol-II and mRNA time series can be done in closed form given fixed values 
of the differential equation parameters. GP inference is based on mean and 
covariance functions. Below 
we describe the GP model of pol-II and mRNA, their 
respective mean functions and covariance functions, and the cross-covariance 
function between pol-II and mRNA. GP inference of the posterior is then a 
standard inference equation which we provide 
for completeness.

The above inference provides a posterior distribution over the profiles $p(t)$ and $m(t)$ given 
known values for the differential 
equation parameters. However, these values are not known and to infer a full posterior 
over both time series and 
these parameters we carry out Markov Chain Monte Carlo (MCMC)
sampling over the parameter values, as described in Section ``Parameter
inference by Hamiltonian Monte Carlo sampling''.

\subsection{GP model of pol-II}
\label{sec:gp-pol2}

For each gene, we model the pol-II activity time series 
in a nonparametric fashion by applying a GP prior over the 
shapes of the time series.
Previous similar GP models~\cite{Lawrence2007,Gao2008,Honkela2010}
have used a squared exponential covariance function for $p(t)$, as
that allows derivation of all the shared covariances in closed form.
This covariance has the limitation that it is stationary, and
functions following it revert to zero away from data.  These
properties severely degrade its performance on our highly unevenly
sampled data.  To avoid this, we model $p(t)$ as an
\emph{integral} of a function having a GP prior with a squared
exponential covariance: then the posterior
mean of $p(t)$ tends to remain constant between observed data.
That is, we model
\begin{equation}
p(t)=p_0 + \int_{u=0}^t v(t) \mathrm{d}t
\label{eq:differential_equation_pol2}
\end{equation}
where $p_0$ is the initial value at time $t=0\mathrm{\;min}$, and assign a GP prior
with the squared exponential covariance over
$v(t)$; as a result $p(t)$ will also have a GP prior whose
covariance function is an integral of the covariance function of
$v(t)$. For mathematical convenience we assume $p(t)=0\mathrm{\;min}$ for $t<0\mathrm{\;min}$, and
set the initial observation time $t_0$ to a sufficiently large value to avoid any
discontinuity resulting from assumption in pol-II or mRNA modeling.

To define the GP prior, we first define the mean function of $p(t)$.
Assume that $v(t)$ is drawn from a zero-mean GP prior 
with a squared exponential covariance function
$k_v(t,t') = \polmag \cdot \exp(-(t-t')^2/\polwidth^2)$ where 
$\polmag$ is a magnitude parameter and $\polwidth$ is a length scale,
which has been parametrised in a non-standard manner to simplify
the derivations.
Then $E[p(t)]=E[p_0] + \int_{u=t_0}^t E[v(t)] \mathrm{d}t =E[p_0]\equiv\mu_p$
for $t\ge 0\mathrm{\;min}$.

Next we compute the corresponding covariance function for the GP prior
of $p(t)$. We have
\begin{multline}
k_p(t,t') \equiv E[(p(t)-\mu_p)(p(t')-\mu_p)] 
= \int_{s=0}^t \int_{s'=0}^{t'} k_v(s,s') \mathrm{d}s \mathrm{d}s' \\
= \frac{\sqrt{\pi}\polmag\polwidth}{2} \int_{s=0}^t 
\bigg(
\textrm{erf}((t'-s)/\polwidth) 
- \textrm{erf}(-s/\polwidth) 
\bigg) ds 
\end{multline}

The remaining integral over the erf functions can be computed using 
integration by parts. After straightforward manipulation, the integral 
becomes
\begin{multline}
k_p(t,t') 
= \frac{\polmag\sqrt{\pi}\polwidth^2}{2} \bigg(
t_{\polwidth}\text{erf}\left(t_{\polwidth}\right)
+ t'_{\polwidth}\text{erf}\left(t'_{\polwidth}\right)
-(t'_{\polwidth}-t_{\polwidth})\text{erf}\left(t'_{\polwidth}-t_{\polwidth}\right) 
\bigg)  \\
+\frac{\polmag \polwidth^2}{2} \bigg(
\exp\left(-t_{\polwidth}^2\right)
+\exp\left(-(t'_{\polwidth})^{2}\right)
-\exp\left(-(t'_{\polwidth}-t_{\polwidth})^2\right)
-1
\bigg) \;.
\end{multline}
where we denoted $t_{\polwidth} = t/\polwidth$ and 
$t'_{\polwidth} = t'/\polwidth$ for brevity.

The right-hand side is the covariance function $k_p(t,t')$ of the \\
integrated squared-exponential GP prior for pol-II.

\subsection{GP model of mRNA}
\label{sec:gp-mrna}

We model the mRNA abundance in a similar nonparametric fashion as
the pol-II activity. Since the mRNA is related to pol-II through
a differential equation, the GP prior of mRNA can be computed from the
GP prior of pol-II through the differential equation. 
In particular, as shown in Eq. \eqref{eq:mrna_integral_appendix}, the mRNA time
series is an integral of the pol-II time series. Since integration is
a linear operation, the expected mRNA time
series is an integral of the expected pol-II time series; that is, the GP mean function of
mRNA is an integral of the mean function of pol-II, so that
\begin{multline}
\mu_m(t)\equiv E[m(t)]
=m_0 e^{\alpha(t_0-t)} + \\
\frac{\beta_0}{\alpha}\left(1-e^{-\alpha t}\right)
+ \beta e^{-\alpha t}\int_{u=0}^t e^{\alpha u} E[p(u-\Delta)] \mathrm{d}u \\
=m_0 e^{-\alpha (t-t_0)} + 
\frac{\beta_0}{\alpha}\left(1-e^{-\alpha t}\right) 
+ \beta e^{-\alpha t}\int_{u=\Delta}^{t} e^{\alpha u} \mu_p \mathrm{d}u \\
=m_0 e^{-\alpha (t-t_0)} + 
\frac{\beta_0}{\alpha}\left(1-e^{-\alpha t}\right) 
+ \frac{\beta \mu_p}{\alpha} (1-e^{-\alpha (t-\Delta)}) 
\end{multline}
where the third line follows since pol-II activity starts at $t=0\mathrm{\;min}$.
Note that the start of pol-II activity at $t=0\mathrm{\;min}$ is for mathematical convenience, and the
initial observation time $t_0$ will be set to a sufficiently large
value so that the $t-\Delta\ge 0\mathrm{\;min}$ for all $t\ge t_0$ and hence
observed mRNA values are integrated over active pol-II only regardless
of delay $\Delta$.

We next compute the corresponding covariance function for the GP prior
of mRNA. The covariance function arises from computing the integral 
relating mRNA to pol-II as follows:
\begin{multline}
k_m(t,t')
\equiv E[(m(t)-\mu_m(t))(m(t')-\mu_m(t'))] \\
= E\bigg[ \bigg( 
\beta e^{-\alpha t}\int_{u=0}^t e^{\alpha u} p(u-\Delta)
\mathrm{d}u 
- \frac{\beta \mu_p}{\alpha} (1-e^{-\alpha (t-\Delta)})
\bigg) \\
\bigg(  
\beta e^{-\alpha t'}\int_{u'=0}^{t'} e^{\alpha u'} p(u'-\Delta)
\mathrm{d}u'
- \frac{\beta \mu_p}{\alpha} (1-e^{-\alpha (t'-\Delta)})
 \bigg)\bigg] \\
= \beta^2 E\bigg[ \bigg( 
e^{-\alpha t}\int_{u=\Delta}^t e^{\alpha u} p(u-\Delta)
\mathrm{d}u 
- \frac{\mu_p}{\alpha} (1-e^{-\alpha (t-\Delta)})
\bigg) \\
\bigg(  
e^{-\alpha t'}\int_{u'=\Delta}^{t'} e^{\alpha u'} p(u'-\Delta)
\mathrm{d}u'
- \frac{\mu_p}{\alpha} (1-e^{-\alpha (t'-\Delta)})
 \bigg)\bigg] 
\end{multline}
where the last equality follows since pol-II activity starts at time 0~min.
The computation of the integrals follows similar steps as computation of the
pol-II GP covariance. The result is 
\begin{equation}
k_m(t,t')
= k_{m,1}(t,t')+k_{m,2}(t,t')+k_{m,3}(t,t')+k_{m,4}(t,t')
\end{equation}
where we divided the covariance function into four parts. The first
part is 
\begin{multline}
k_{m,1}(t,t') 
= \frac{\sqrt{\pi} \polwidth \polmag \rnamag}{2\alpha^2}
\bigg(  
      \bigg(t_{\Delta} - \frac{1}{\alpha} + \frac{\exp(-\alpha t'_{\Delta})}{\alpha}\bigg) \mathrm{erf}\bigg(\frac{t_{\Delta}}{\polwidth}\bigg)\\
  +  \bigg(t'_{\Delta} - \frac{1}{\alpha} + \frac{\exp(-\alpha t_{\Delta})}{\alpha}\bigg) \mathrm{erf}\bigg(\frac{t'_{\Delta}}{\polwidth}\bigg)
  - (t_{\Delta}-t'_{\Delta})\mathrm{erf}\bigg(\frac{t_{\Delta}-t'_{\Delta}}{\polwidth}\bigg) \bigg)  
\end{multline}
where $t_{\Delta}=\max(0\mathrm{\;min}, t-\Delta)$.
The second part is
\begin{multline}
k_{m,2}(t,t')
 = \frac{\polwidth^2 \polmag \rnamag}{2 \alpha^2}
\bigg(
\exp\bigg(-\bigg(\frac{t_{\Delta}}{\polwidth}\bigg)^2\bigg) + \exp\bigg(-\bigg(\frac{t'_{\Delta}}{\polwidth}\bigg)^2\bigg) \\
- \exp\bigg(-\bigg(\frac{t_{\Delta}-t'_{\Delta}}{\polwidth}\bigg)^2\bigg) - 1  \bigg) \;.
\end{multline}
The third part is
\begin{multline}
k_{m,3}(t,t') 
= -\sqrt{\pi}\polwidth \polmag \rnamag 
\bigg(
\frac{\alpha^{-3}}{4}\exp(\alpha^2\polwidth^2/4 + \alpha(t_{\Delta}-t'_{\Delta})) \\
(\mathrm{erf}(\alpha\polwidth/2+t_{\Delta}/\polwidth) - \mathrm{erf}\big(\alpha\polwidth/2+(t_{\Delta}-t'_{\Delta})/\polwidth)\big)\\
+ \frac{\alpha^{-3}}{4}\exp(\alpha^2 \polwidth^2/4 - \alpha t'_{\Delta} -\alpha t_{\Delta})
\big(\mathrm{erf}(\alpha\polwidth/2) - \mathrm{erf}(\alpha\polwidth/2-t'_{\Delta}/\polwidth)\big)\\
- \frac{\alpha^{-3}}{2}\exp(\alpha^2 \polwidth^2/4 - \alpha t'_{\Delta}) 
\big(\mathrm{erf}(\alpha\polwidth/2) - \mathrm{erf}(\alpha\polwidth/2-t'_{\Delta}/\polwidth)\big)
\bigg)
 \;.
\end{multline}
The fourth part is
\begin{multline}
k_{m,4}(t,t') 
= -\sqrt{\pi}\polwidth \polmag \rnamag 
\bigg(
\frac{\alpha^{-3}}{4}\exp(\alpha^2 \polwidth^2/4 
- \alpha(t_{\Delta}-t'_{\Delta})) \\
\big(\mathrm{erf}(\alpha\polwidth/2+t'_{\Delta}/\polwidth) -
\mathrm{erf}(\alpha\polwidth/2-(t_{\Delta}-t'_{\Delta})/\polwidth)\big) \\
+ \frac{\alpha^{-3}}{4}\exp(\alpha^2 \polwidth^2/4 - \alpha t_{\Delta} -\alpha t'_{\Delta})
\big(\mathrm{erf}(\alpha\polwidth/2) - \mathrm{erf}(\alpha\polwidth/2-t_{\Delta}/\polwidth)\big)\\
- \frac{\alpha^{-3}}{2}\exp(\alpha^2 \polwidth^2/4 - \alpha t_{\Delta})
\big(\mathrm{erf}(\alpha\polwidth/2) - \mathrm{erf}(\alpha\polwidth/2-t_{\Delta}/\polwidth)\big)
\bigg) \;.
\end{multline}

\subsection{GP joint model}
\label{sec:gp-cross}
To define the full GP prior over both pol-II and mRNA, it remains to
define the cross-covariance function between pol-II and mRNA. The full
GP covariance is defined by the individual covariances of pol-II
and mRNA and the cross-covariance.

The cross-covariance function between (noiseless) mRNA abundance
$m(t)$ at time $t$ and (noiseless) pol-II activity $p(t')$ at
time $t'$ is computed with similar steps as the computation of the 
mRNA covariance function. The result is
\begin{multline}
k_{mp}(t,t') 
= E[(m(t)-\mu_m(t))(p(t')-\mu_p(t'))] \\
= k_{mp,1}(t,t')+k_{mp,2}(t,t')+k_{mp,3}(t,t') 
\end{multline}
where for convenience we separated the kernel function into a sum of
three components. 
The first component part of the kernel can be written as
\begin{multline}
k_{mp,1}(t,t') = \\
-\frac{\sqrt{\pi} \beta^2\sqrt{\polmag} \polwidth}{2 \alpha^2} 
\exp\bigg( \bigg(\frac{\alpha \polwidth}{2} \bigg)^2 - \alpha t_{\Delta} +\alpha t' 
         \bigg)\\
\cdot \bigg[\mathrm{erf}\bigg(\frac{\alpha \polwidth}{2} + \frac{t'}{\polwidth} \bigg)
          -\mathrm{erf}\bigg(\frac{\alpha \polwidth}{2}  + \frac{t'-t_{\Delta}}{\polwidth} \bigg)
       \bigg]\\
-\frac{\sqrt{\pi} \beta^2\sqrt{\polmag} \polwidth}{2 \alpha^2} 
\exp\bigg(\bigg(\frac{\alpha \polwidth}{2} \bigg)^2-\alpha t_{\Delta}
\bigg)\\
\cdot\bigg[\mathrm{erf}\bigg(\frac{\alpha \polwidth}{2} -\frac{t_{\Delta}}{\polwidth} \bigg)
-\mathrm{erf}\bigg(\frac{\alpha \polwidth}{2} \bigg)
     \bigg] \;.
\end{multline}
The second component can be written as
\begin{multline}
k_{mp,2}(t,t') = \\
  - \frac{\beta^2\sqrt{\polmag} \polwidth^2}{2 \alpha} \bigg[\exp\bigg(-\bigg(\frac{t_{\Delta}-t'}{\polwidth}\bigg)^2\bigg)-\exp\bigg(-\bigg(\frac{t_{\Delta}}{\polwidth}\bigg)^2\bigg)\\
+1-\exp\bigg(-\bigg(\frac{t'}{\polwidth}\bigg)^2\bigg)\bigg] \;.
\end{multline}
The third component can be written as
\begin{multline}
k_{mp,3}(t,t') = -\frac{\sqrt{\pi} \beta^2\sqrt{\polmag} \polwidth}{2 \alpha}
    \bigg [ (t_{\Delta}-t'-1/\alpha) \mathrm{erf} ((t_{\Delta}-t')/\polwidth)\\
      -(t_{\Delta}-1/\alpha) \mathrm{erf} (t_{\Delta}/\polwidth)
      -(t'+\exp(-\alpha t_{\Delta})/\alpha) \mathrm{erf} (t'/\polwidth) \bigg] \;.
\end{multline}

\subsection{Observation model}
In order to fit the models of the pol-II and mRNA functions to
observations, we need an observation model.
It is assumed that we observe noisy values $\tilde{m}(t) = m(t) + e_m(t)$ and $\tilde{p}(t) = p(t) +
e_p(t)$ where $e_m(t)$ and $e_p(t)$ are zero-mean Gaussian noise
independently sampled for each time point. For simplicity we assume
the noise variance of $e_p(t)$ is a constant $\polnoisevar$ and infer
it as a parameter of the model. We estimate the mRNA noise variances $\rnanoisevar(t)$
for each time point $t$ 
as sums of a shared constant $\sigma_{m}^2$ and
a fixed variance inferred by BitSeq by combining the technical quantification
uncertainty from BitSeq expression estimation with an estimate of
biological variance from the BitSeq differential expression model
(full details are in Sec. RNA-seq data processing).

Since the noise is zero-mean, the GP prior for the noisy observations 
has the same means as the noiseless means, that is,  
$E[\tilde{m}(t)]=E[m(t)]$ and
$E[\tilde{p}(t)]=E[p(t)]$. Since the noise is independently added to
each observation, the covariance function of observed pol-II becomes
\begin{equation}
k_{\tilde{p}}(t,t') = k_{p}(t,t') + \delta(t,t') \polnoisevar
\end{equation}
where $\delta(t,t')=1$ if $t=t'$ and zero otherwise,
the covariance function of observed mRNA becomes
\begin{equation}
k_{\tilde{m}}(t,t') = k_{m}(t,t') + \delta(t,t') \rnanoisevar(t) \;,
\end{equation}
and the 
cross-covariance function between observed
pol-II and mRNA is the same as the noiseless version so that
\begin{equation}
k_{\tilde{m}\tilde{p}}(t,t') = k_{mp}(t,t') \;.
\end{equation}

The GP prior over time series functions and the observation model
together define a full probability model for the pol-II and mRNA
data. As the observations are noisy and available only at a small set
of time points, we will apply Bayesian inference to infer the underlying
time series $m(t)$ and $p(t)$ from the observations.

\subsection{Covariance matrix for GP inference}

Given a set of time points, here
the 10 time points
\begin{align*}
  T_{obs} &= t_0 + (0,5,10,20,40,80,160,320,640,1280) \\
    &= (t_1, \dots, t_N)
\end{align*}
where $t_0$ is the initial observation time and the numbers denote time in minutes,
and the corresponding observation data consisting of
$N=10$ pol-II observations and $N=10$ mRNA observations 
$\alldata = (\tilde{p}(t_1),\ldots,\tilde{p}(t_N),\tilde{m}(t_1),\ldots,\tilde{m}(t_N))$,
we wish to compute the posterior distribution of
GP hyperparameters, and to predict the shape of
the underlying time series functions $p(t)$ and $m(t)$ given the posterior.
We will especially wish to study delay between pol-II and mRNA; 
for mathematical convenience we set $t_0=300\mathrm{\;min}$ and consider mRNA delay
parameters $0\le\Delta\le 300\mathrm{\;min}$.

For GP inference, given the hyperparameters we must compute the prior
GP covariance matrix for the observations $\alldata$. We describe the
matrix here in a general form which is needed later for inference of
time series values at previously unseen time points.

The covariance matrix describes covariance between measurements
at one set of time points (indexed by rows of the matrix) and another
set of time points (indexed by columns of the matrix). 
Let $T_{row}=(t_{row,1},\ldots,t_{row,N_{row}})$ be a vector of
$N_{row}$ time indices for rows of the matrix, and let 
$T_{col}=(t_{col,1},\ldots,t_{col,N_{col}})$ be the vector of 
$N_{row}$ time indices for columns of the matrix.

The resulting
covariance matrix $K(T_{row},T_{col})$ has the block structure
\begin{equation}
K(T_{row},T_{col}) =
\bigg[
\begin{array}{cc}
K_{\tilde{p}} & K_{\tilde{p}\tilde{m}} \\
K_{\tilde{m}\tilde{p}} & K_{\tilde{m}}
\end{array}
\bigg] 
\end{equation}
where each block is a $N_{row}\times N_{col}$ matrix of covariance function values
between the time points $t\in T_{row}$ and the
time points $t'\in T_{col}$, so that 
$K_{\tilde{p}}$ is composed of values $k_{\tilde{p}}(t,t')$,
$K_{\tilde{m}}$ is composed of values $k_{\tilde{m}}(t,t')$,
$K_{\tilde{m}\tilde{p}}$ is composed of the cross-covariance values
$k_{\tilde{m}\tilde{p}}(t,t')$,
and $K_{\tilde{p}\tilde{m}}$ is composed of the cross-covariance values
$k_{\tilde{m}\tilde{p}}(t',t)$.
The covariance matrix of observed data is then simply
$K_{obs}=K(T_{obs},T_{obs})$.

\subsection{Marginal likelihood function}
The analytical tractability of the GP model allows us to marginalise
out the latent functions $p(t)$ and $m(t)$ to compute a marginal
likelihood that only depends on the model parameters.
The marginal probability density of the observations $\alldata$ is
Gaussian and the marginal log-likelihood is
\begin{equation}
\log
P(\alldata) 
= (1/2)(-d \log(2\pi) - \log(|K_{obs}|) - u^\top K_{obs}^{-1} u)
\end{equation}
where $P$ denotes the marginal probability density,
$d=20$ is the total number of pol-II and mRNA observations and 
$u$ is the column vector of observations with their expected values subtracted,
$u=\big[\tilde{p}(t_1)-E[\tilde{p}(t_1)],\ldots,\tilde{p}(t_N)-E[\tilde{p}(t_N)],\tilde{m}(t_1)-E[\tilde{m}(t_1)],\ldots,\tilde{m}(t_N)-E[\tilde{m}(t_N)]\big]^\top$.

\subsection{Posterior prediction}
The analytical tractability of the GP model also allows us to
obtain the full posterior distribution over the latent functions in
closed form given the parameters.
Given the observed data, we can thus compute the mean and
covariance of the underlying time series function values at each time
point, as expectations over the posterior distribution of the
underlying functions. 
For $N^*$ new time points $T^*=(t^*_1,\ldots,t^*_{N^*})$
the posterior mean is
\begin{equation}
E[[\tilde{p}(t^*_{1}),\ldots,\tilde{p}(t^*_{N^*}),\tilde{m}(t^*_{1}),\ldots,\tilde{m}(t^*_{N^*})]|\alldata]
= u^*_{\textrm{prior}} + K(T^*,T_{obs}) K_{obs}^{-1} u
\end{equation}
where 
\begin{equation}
u^*_{\textrm{prior}}=\big[E[\tilde{p}(t^*_{1})],\ldots,E[\tilde{p}(t^*_{N^*})],
E[\tilde{m}(t^*_{1})],\ldots,E[\tilde{m}(t^*_{N^*})]\big]^\top
\end{equation}
is the vector of prior means computed at the new time points, and
the posterior covariance matrix is
\begin{multline}
Cov[(\tilde{p}(t^*_{1}),\ldots,\tilde{p}(t^*_{N^*}),\tilde{m}(t^*_{1}),\ldots,\tilde{m}(t^*_{N^*}))|\alldata]\\
= K(T^*,T^*) - K(T^*,T_{obs}) K_{obs}^{-1} K(T^*,T_{obs})^{\top}\;.
\end{multline}

The log-likelihood and predictions of function values described here
are computed given fixed values of hyperparameters of the GP prior and
the observation model. We will compute a posterior distribution for
the hyperparameters, given suitable prior distributions for each. This
will allow summarisation of underlying pol-II and mRNA functions and
GP parameters over the posterior distribution of the
hyperparameters. We next describe the prior distributions of
hyperparameters and then describe the sampling based inference of
hyperparameter posterior distributions.

\subsection{Parameter prior distributions}
\label{sec:priors}

All parameters except the delay $\Delta$ have approximately uniform
bounded logistic-normal priors.  These priors were used because of
convenience: they allow easy Hamiltonian Monte Carlo sampling that
requires very little tuning (see below for details).

The density of the logistic-normal prior $\mathrm{logit}\mbox{-}\mathrm{normal}(\mu,
\sigma^2, a, b)$ with location parameter $\mu$ and scale parameter
$\sigma^2$ for variable $\theta$ bounded to the interval $[a,b]$ is
\begin{multline}
  \label{eq:logitnormal_pdf}
  p(\theta | \mu, \sigma^2, a, b) = \\
  \frac{1}{\sqrt{2 \pi \sigma^2}}
  \exp\left( -\frac{(\mathrm{logit}((\theta - a)/(b-a)) - \mu)^2}{2 \sigma^2}  \right)
  \cdot \frac{b - a}{(\theta - a)(b - \theta)},
\end{multline}
where $\mathrm{logit}(x) = \log(x / (1-x))$. We use $\mu=0$ and
$\sigma = 2$ which lead to an approximately uniform distribution on
the interval $[a,b]$.  The interval bounds $a, b$ for all variables
are presented in Table \ref{tab:parameter_values}.

For the delay $\Delta$ we use a prior with $\mu=-2, \sigma=2$ to
reflect our prior belief that the delays should in most cases be
small.
For $\beta$ and $\polwidth$ we set the 
priors with respect to $\beta^2$ and 
$2/\polwidth^2$ respectively, because these are more convenient
as model parameters.

\begin{table}[htb]
\centering
\begin{tabular}{ccc}
\hline
 Parameter & Lower bound $a$ & Upper bound $b$ \\
\hline
$2/\polwidth^2$
&  $(1280 \mathrm{\;min})^{-2}$ & $(5 \mathrm{\;min})^{-2}$ \\ 
&  $\approx 6.1\cdot 10^{-7}\mathrm{\;min}^{-2}$ & $= 4\cdot 10^{-2}\mathrm{\;min}^{-2}$ \\
$\polmag$ & $2\cdot 10^{-4}$ $\hat{\sigma}^2_{Pol2}$ &  $\hat{\sigma}^2_{Pol2}$ \\
$\polnoisevar$ & $0.05 \hat{\sigma}^2_{Pol2}$ & $\hat{\sigma}^2_{Pol2}$ \\
$\alpha$ & $1\cdot 10^{-6}\mathrm{\;min}^{-1}$& $\log(2) \mathrm{\;min}^{-1}\approx 0.69 \mathrm{\;min}^{-1}$ \\
$\beta^2$ & $1\cdot 10^{-6}\mathrm{\;min}^{-2}$ & $1 \mathrm{\;min}^{-2}$\\
$\Delta$ & $0 \mathrm{\;min}$ & $299 \mathrm{\;min}$ \\ 
$\beta_0$ & $0 \mathrm{\;min}^{-1}$ & $1 \mathrm{\;min}^{-1}$ \\
$m_0$ & $0$ & $2$ \\
$\mu_p$ & $0$ & $1$ \\
\hline
\end{tabular}
\caption{Bounds for bounded logistic-normal priors of differential
  equation parameters in the GP inference of pol-II and
  mRNA time series. Each parameter is bounded to an interval $[a,b]$,
  we list the values of the lower bound $a$ and upper bound $b$.
Here $\hat{\sigma}^2_{Pol2}$ is the empirical variance of the pol-II
time series after preprocessing.
\label{tab:parameter_values}}
\end{table}

\subsection{Parameter inference by Hamiltonian Monte Carlo sampling}
\label{sec:mcmc}

Given the data and the priors for the parameters, we apply fully
Bayesian inference with Hamiltonian Monte Carlo (HMC) sampling~\cite{Duane1987} to
obtain samples from the posterior distribution of the parameters.  HMC
is a MCMC algorithm that uses gradients of
the target distribution to simulate a Hamiltonian dynamical system
with an energy function based on the target distribution.  This allows
taking long steps while maintaining a high acceptance rate in the
sampling.

In order to apply HMC more easily, we transform all parameters to an
unbounded space using the logistic transformation.  The
logistic-normal priors correspond to normal priors on the transformed
variables, which effectively prevent the sampler from wandering off to
the saturated region of the transformation near the bounds of the
intervals.

We run 4 parallel chains starting from different random initial states
to allow convergence checking.  We use the HMC implementation from
NETLAB toolkit in Matlab with momentum persistence and number of leap
frog steps $\tau = 20$ which were found to work well in all cases.
The step length $\epsilon$ is tuned separately for every model (see
below).  After tuning, each chain is run for 10000 iterations.  The
samples are then thinned by a factor of 10, and the first half of the
samples are discarded, leaving 500 samples from each chain, 2000 in
all.  Convergence is monitored using the potential scale reduction
factor $\sqrt{\hat{R}}$~\cite{Gelman1992}.  $\sqrt{\hat{R}}$ is
computed separately for each variable, and if any of them is greater
than 1.2, the result is discarded and a new sample obtained in a
similar manner.  The 9 genes that did not converge after 10 iterations of
this process were removed from further analysis.  In most cases these had severely
multimodal delay distributions that were difficult to sample from and
would have made further analysis difficult.

\subsubsection{Tuning} 
The applied logistic transformation and priors together allow using
the same global step length $\epsilon$ for all variables, or using the
identity matrix as the mass matrix in the HMC formulation.  The step
length $\epsilon$ was determined by trying different alternatives in
the set $\{10^{-5}, 10^{-4}, 10^{-3}, 0.003, 0.005, 0.01, 0.03, 0.05,
0.07, 0.1,$ $0.3, 0.5, 1\}$ in increasing order, running the sampler for
100 steps and using the largest value with at least $80\%$ acceptance
rate.  This target rate is higher than usual in random walk MCMC
because HMC acceptance rate should be nearly $100\%$ even with very
long steps if the Hamiltonian system is simulated accurately.

\subsubsection{Summarisation of inference results}

The inference results are summarised using the median of the posterior
samples.  This is a convenient statistic because it is invariant to
transformations of the parameter space, such as those used during the
sampling.

\subsection{Validation of the GP modelling results}

In order to validate the GP model, we implemented inference for the
same ODE using a smoothing spline fit for pol-II.  A comparison of the
results for the subset of genes that yielded reliable results with the
spline approach is presented in Fig.~\ref{fig:splinecomparison}.

\section{Filtering of results}

Reliable posterior samples were obtained for models of 4373 genes.
4304 of these had multiple-exon transcripts, and could thus be used
for intron analyses.  These genes were further filtered to remove
bad fits by only keeping genes that satisfy the following:
\begin{enumerate}
\item The global maximum $t_{\text{max}}$ of $p(t)$ posterior mean
  $\bar{p}(t)$ in the interval $t \in [0 \text{ min}, 1280 \text{
    min}]$ occurs in the interval $t_{\text{max}} \in (1 \text{ min},
  160 \text{ min})$.  This condition ensures the profile has a peak in
  the densely sampled region which is necessary for accurate
  estimation of the delay.
\item The posterior median delay $\hat{\Delta} < 120 \text{ min}$.
  Because of the increasingly sparse sampling, longer delay estimates
  were considered unreliable.  The specific cut-off was determined by
  visual inspection of the fits to rule out implausible ones.
\item The posterior mean $\bar{p}(t)$ of $p(t)$ does not change too
  much just before $t=0 \text{ min}$.  This condition is necessary to
  avoid cases where a long delay pushes distinctive features of $m(t)$
  to $p(t), t < 0 \text{ min}$, which conflicts with the assumption
  that the system is at a steady state until $t = 0 \text{ min}$.
  Quantitatively, we define an index
  \begin{equation}
    \label{eq:begdev}
    D = D_- - D_+ = D_{[-30\mathrm{\;min},0\mathrm{\;min}]} - D_{[0\mathrm{\;min}, 10\mathrm{\;min}]}
  \end{equation}
  where
  \begin{equation}
    \label{eq:begdev2}
    D_{I} = \left(\max\limits_{t \in I}[\bar{p}(t)] - \min\limits_{t \in I}[\bar{p}(t)] \right) / \max\limits_{t \in [-30\mathrm{\;min}, 1280\mathrm{\;min}]}[\bar{p}(t)],
  \end{equation}
  and only include genes with
  \begin{equation}
    \label{eq:begdev_rule}
    D < 0.05.
  \end{equation}
  Intuitively, $D_{I}$ looks at the magnitude of change in
  $\bar{p}(t)$ in the interval $I$ relative to the global magnitude of
  change in $\bar{p}(t)$.  The final statistic $D$ looks for genes
  that have small changes in $[-30\mathrm{\;min}, 0\mathrm{\;min}]$, but forgives genes with early
  large changes in $[0\mathrm{\;min}, 10\mathrm{\;min}]$ because these would often spill over to
  $t < 0\mathrm{\;min}$ because of the properties of the GP model.  The cut-off
  $0.05$ represents $5\%$ change in magnitude, which seems reasonably
  small.  The main conclusions of the work are robust to different cutoffs,
  as demonstrated in Fig.~\ref{fig:delay_tails2} below.
\end{enumerate}
After these filtering steps, there were 1814 genes left for the
analysis.

Main results under an additional filter of setting a maximum for
posterior inter-quartile range are presented in
Fig.~\ref{fig:iqrboundplots}.

\section{Synthetic data generation}

The synthetic data were generated by fitting a GP with the MLP
covariance~\cite{gpml06} to the Pol-II measurements of the gene TIPARP
(ENSG00000163659), and numerically solving the mRNA level using
Eq.~\eqref{eq:mrna_integral_appendix} with the GP posterior mean as $p(t)$.
The parameters used were: $\Delta \in \{0, 10, 20, 30\}\mathrm{\;min}$, $t_{1/2} =
\log(2) / \alpha \in \{2, 4, 8, 16, 32, 64\}\mathrm{\;min}$, $\beta_0 = 0.005$,
$\beta = 0.03$, $m_0 = 0.008/\alpha$.  The parameter values were
chosen empirically to get profiles that approximately fitted the
actual mRNA observations while looking reasonable and informative
across the entire range of parameter values.

\section{Supplementary Results}
\label{sec:suppl-results}

\begin{figure}[tp]
  \centering
  \includegraphics{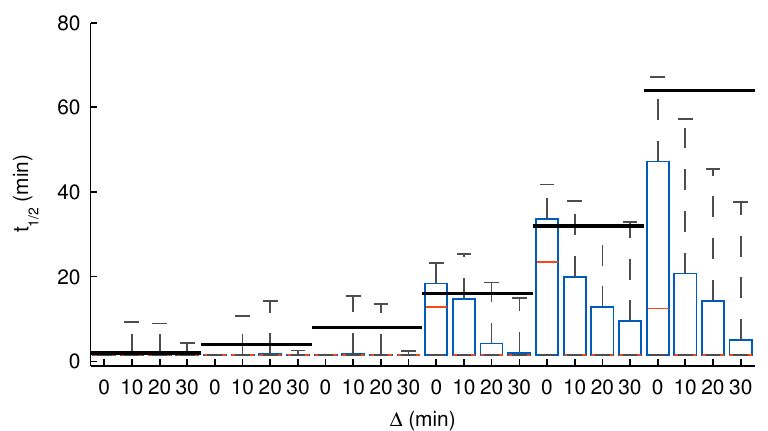}
  \caption{Boxplots of parameter posterior distributions illustrating
    parameter estimation performance on synthetic data for the mRNA
    half life $t_{1/2} = \log(2)/\alpha$.  The strong
    black lines indicate the ground truth used in data generation.
    The box extends from 25th to 75th percentile of the posterior distribution while the whiskers
    extend from 9th to 91st percentile.
    The model often underestimates the half lives, especially
    in the presence of a significant delay.
\label{fig:synthetic_halflives}
}
\end{figure}

\begin{figure*}[t]
  \centering
  \includegraphics{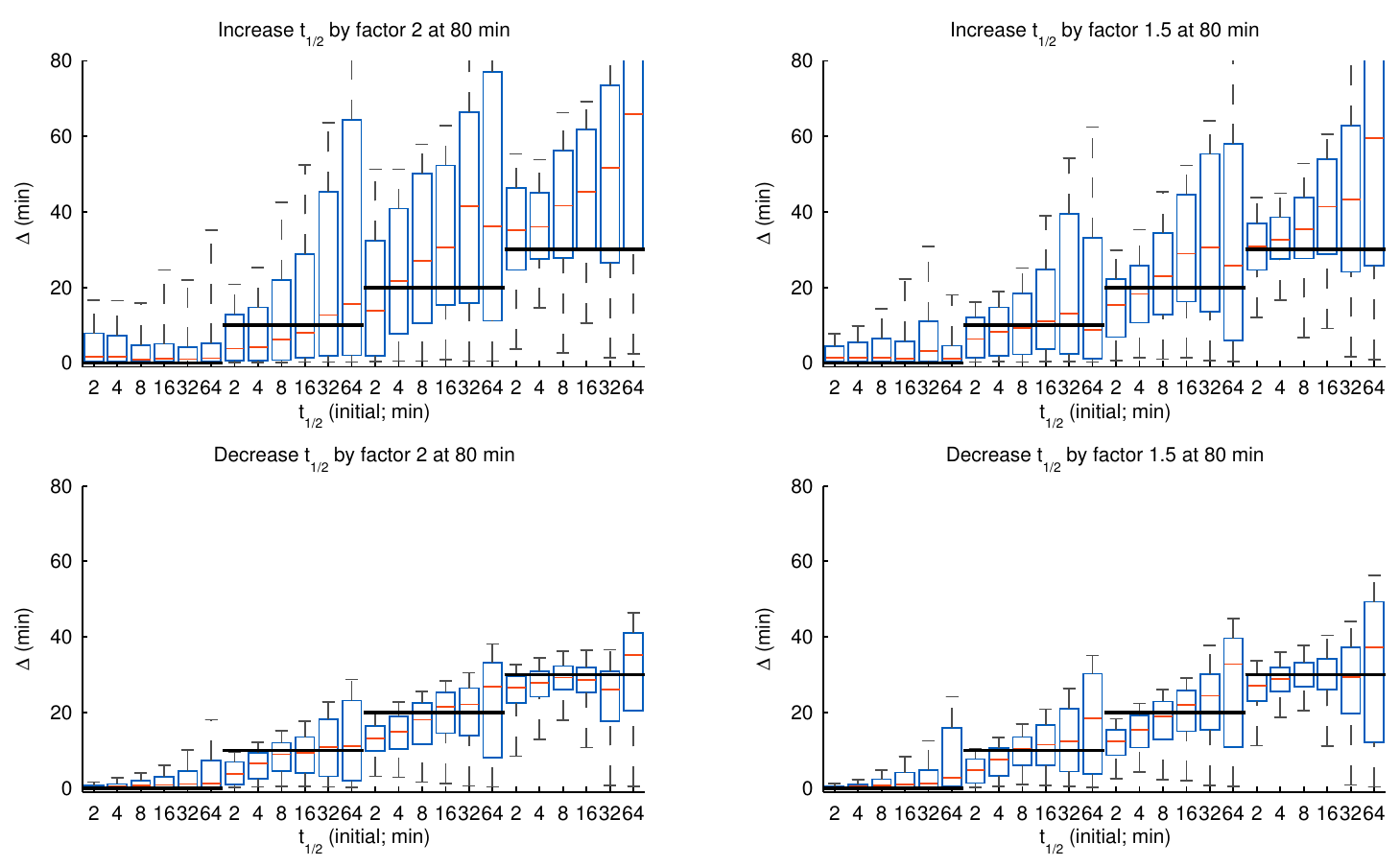}
  \caption{Boxplots of parameter posterior distributions illustrating
    parameter estimation performance on synthetic data for the delay
    parameter $\Delta$.  The strong
    black lines indicate the ground truth used in data generation.
    The box extends from 25th to 75th percentile of the posterior distribution while the whiskers
    extend from 9th to 91st percentile.
    This is a counterpart of Fig.~\ref{fig:synthetic} in a situation
    where the simulated mRNA half-life $t_{1/2}$ changes during the
    time course, something our model cannot capture.
    The simulated changes are point changes up or down with a factor
    of 1.5 or 2 at 80 min.
    The results show that delay estimates remain accurate and reliable,
    with the true value always in the high posterior density region,
    and demonstrate the conservativeness of the estimates with no sign
    of serious overestimation of small delays.
\label{fig:synthetic_changing}
}
\end{figure*}

\begin{figure*}[tp]
  \centering
  \includegraphics{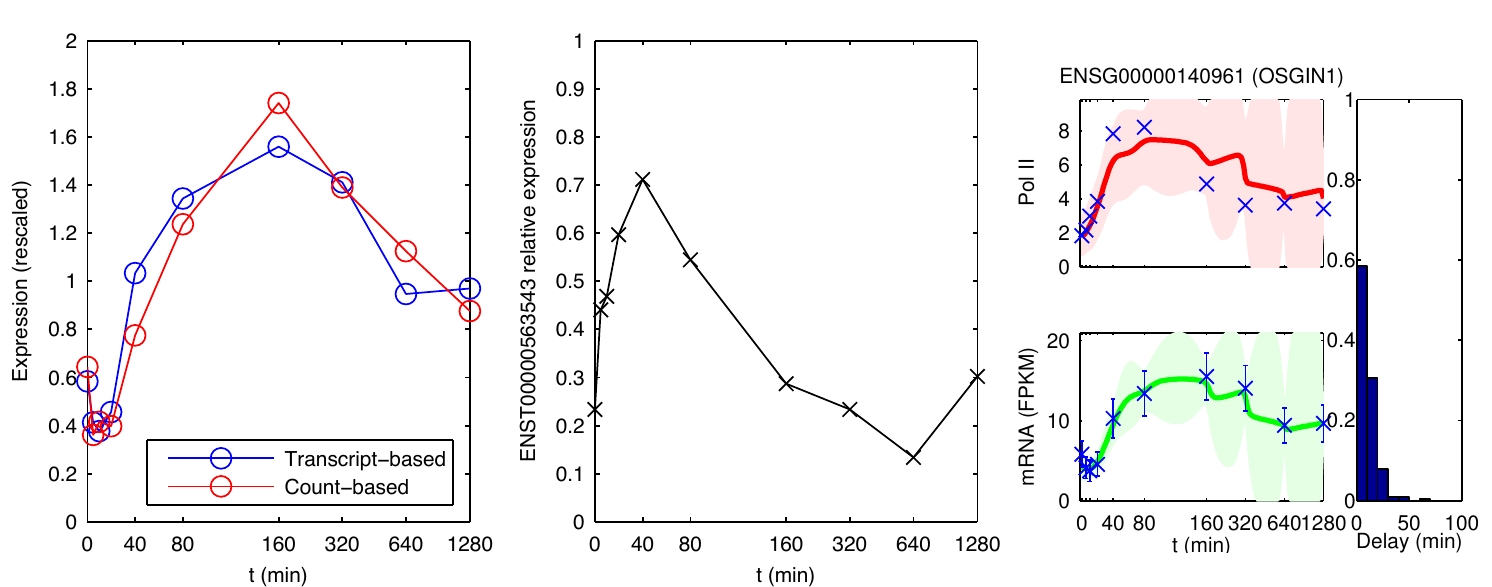}
  \caption{An illustration of how proper summarisation of the
    RNA-seq data is important for ruling out confounding effects from
    changing splicing patterns during the experiment.
    Left: Gene expression time series of the gene OSGIN1 from
    RNA-seq data using either transcript-based RNA-seq data
    summarisation as in the paper or using a simpler summarisation
    of counts of reads aligned uniquely to the mRNA transcripts of
    OSGIN1.
    Centre: Proportion of transcript ENST00000563543 out of all OSGIN1
    transcripts. At 567 bp, this mRNA transcript is much shorter than
    the other major transcripts whose mRNAs are around 2 kb.
    Right: The model fit for OSGIN1 shows no evidence of significant
    delay, while the count-based profile in the left figure
    would suggest a longer delay.
  \label{fig:osgin1_example}}
\end{figure*}

\begin{figure}[tp]
  \centering
  \includegraphics{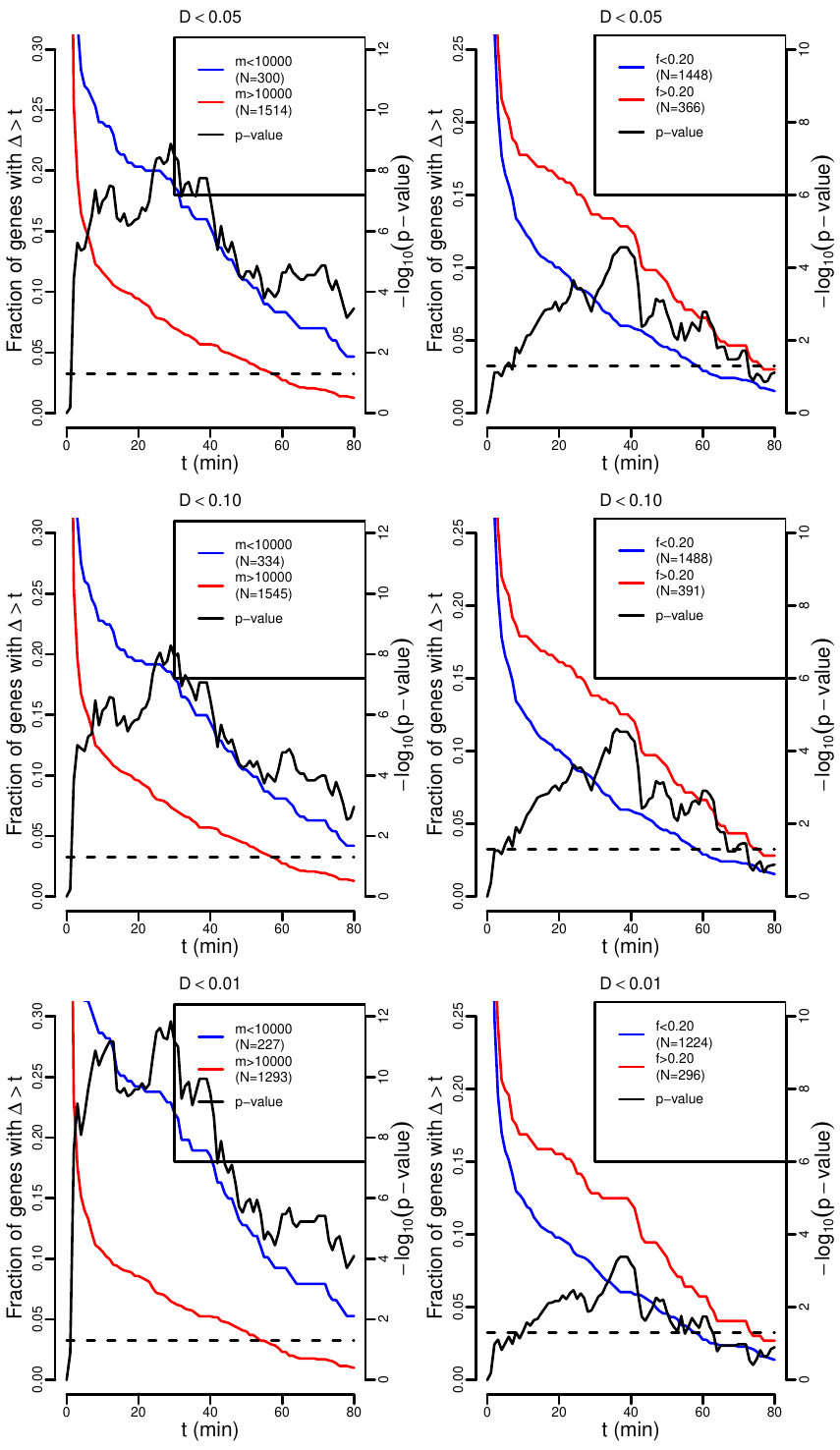}
  \caption{Alternative versions of Fig.~\ref{fig:delay_tails} of the main paper:
    tail probabilities for delays for different cut-offs for $D$
    in Eq.~\eqref{eq:begdev_rule}.  Top: $D < 0.05$ (value used for
    main results), middle: $D < 0.1$, bottom: $D < 0.01$.
    Left: genes whose longest pre-mRNA transcript is short
    ($m$ is the length from transcription start to end).
    Right: genes with relatively long last introns ($f$ is the
    ratio of the length of the last intron of the longest annotated
    transcript of
    the gene divided by the length of that transcript pre-mRNA).
    The fraction of genes with long delays $\Delta$ is shown by
    the red and blue lines (left axis).
    In both subplots, the black curve denotes the $p$-values of
    Fisher's exact test conducted separately at each point (right
    axis) with the dashed line denoting $p<0.05$ significance
    threshold.
    The general shapes of the curves are the same in every case.
\label{fig:delay_tails1}}
\end{figure}

\begin{figure}[tp]
  \centering
  \includegraphics{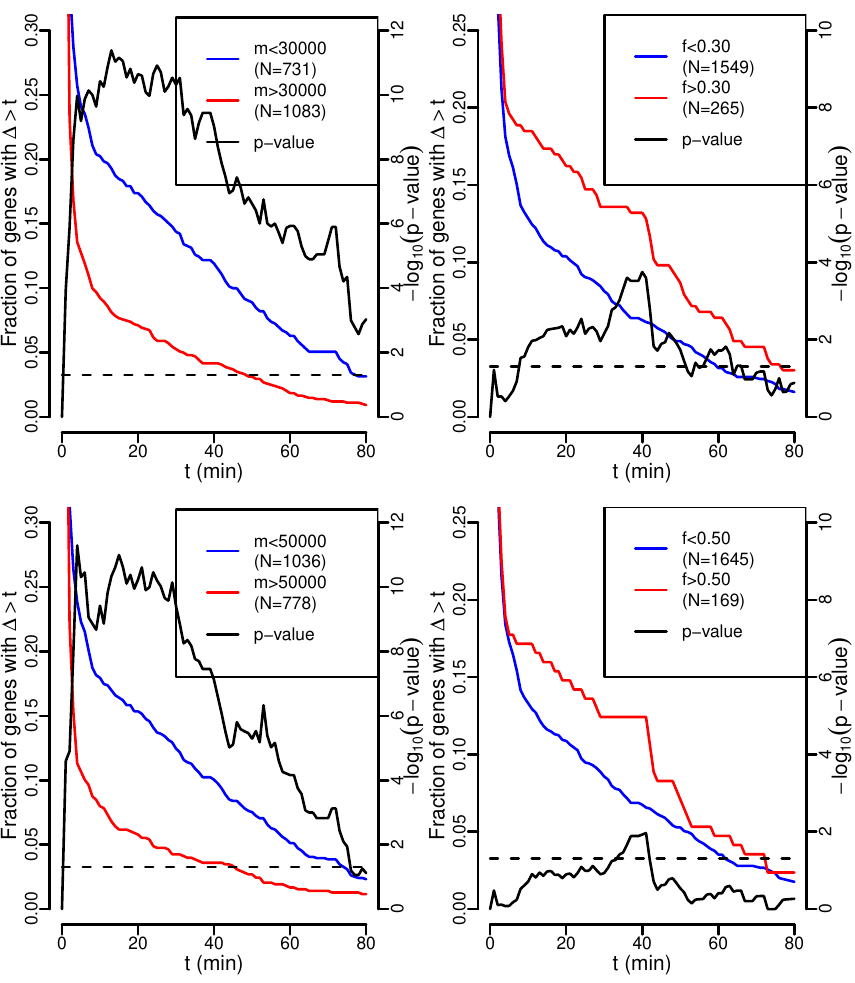}
  \caption{Alternative versions of Fig.~\ref{fig:delay_tails} of the main paper:
    different cut-offs for $f$ and $m$.
    Left: genes whose longest pre-mRNA transcript is short
    ($m$ is the length from transcription start to end).
    Right: genes with relatively long last introns ($f$ is the
    ratio of the length of the last intron of the longest annotated
    transcript of
    the gene divided by the length of that transcript pre-mRNA).
    The fraction of genes with long delays $\Delta$ is shown by
    the red and blue lines (left axis).
    In both subplots, the black curve denotes the $p$-values of
    Fisher's exact test conducted separately at each point (right
    axis) with the dashed line denoting $p<0.05$ significance
    threshold.
    The general shapes of the curves are the same in every case.
\label{fig:delay_tails2}}
\end{figure}

\begin{figure*}[tb]
  \centering
  \includegraphics{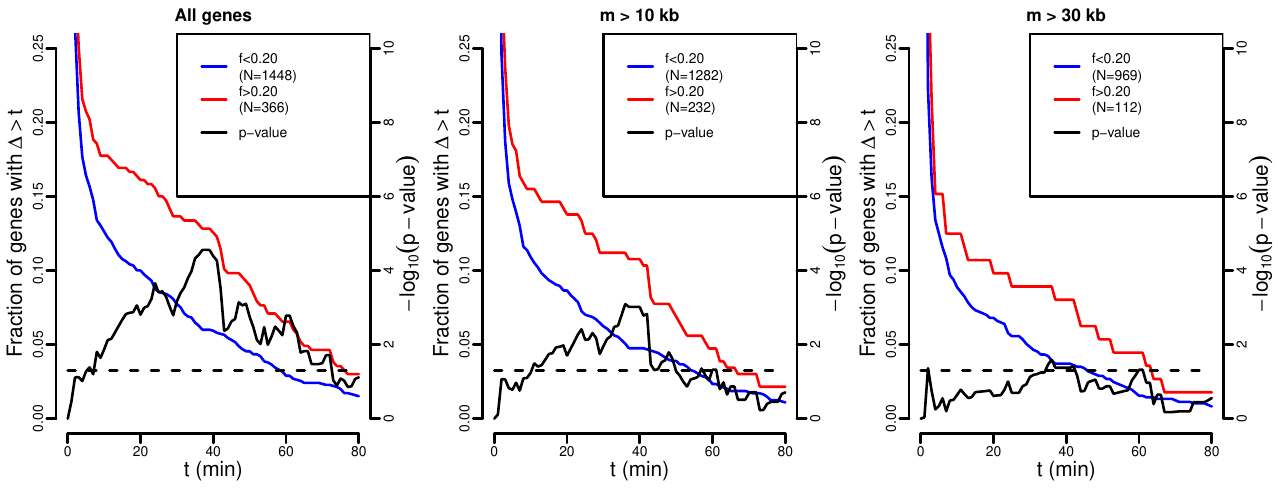}
  \caption{Alternative versions of Fig.~\ref{fig:delay_tails} (right) of the main paper:
    explore the dependence on $f$ for genes with a lower bound on
    mRNA length $m$.
    The fraction of genes with long delays $\Delta$ is shown by
    the red and blue lines (left axis).
    The black curve denotes the $p$-values of
    Fisher's exact test conducted separately at each point (right
    axis) with the dashed line denoting $p<0.05$ significance
    threshold.
    The general shapes of the curves are the same in every case.
    \label{fig:delay_tails_lenfilter}}
\end{figure*}

\begin{figure}[tp]
  \centering
  \includegraphics{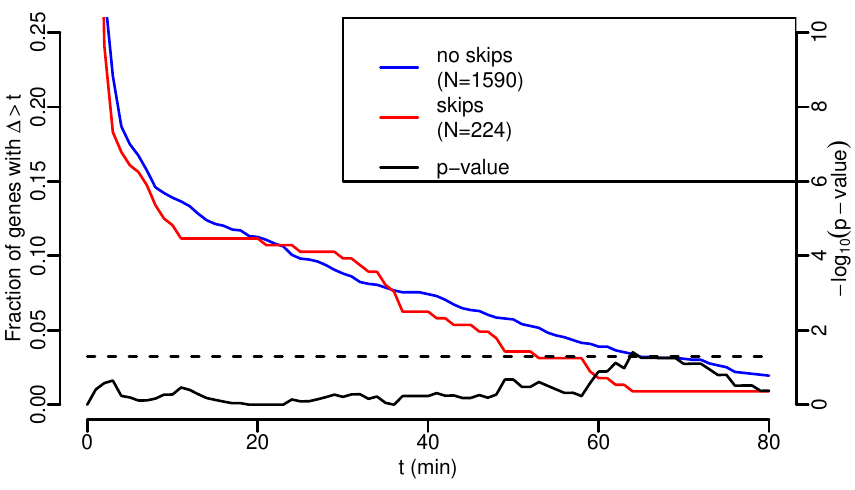}
  \caption{Proportion of long posterior median delays for genes
    with/without annotated exon skipping.
    The fraction of genes with long delays $\Delta$ is shown by
    the red (no skipped exons) and blue (skipped exons) lines (left axis).
    The black curve denotes the $p$-values of
    Fisher's exact test conducted separately at each point (right
    axis) with the dashed line denoting $p<0.05$ significance
    threshold.
    There is no clear difference between the two groups.
    \label{fig:delay_tails_exonskip}}
\end{figure}

\begin{figure}[tp]
  \centering
  \includegraphics{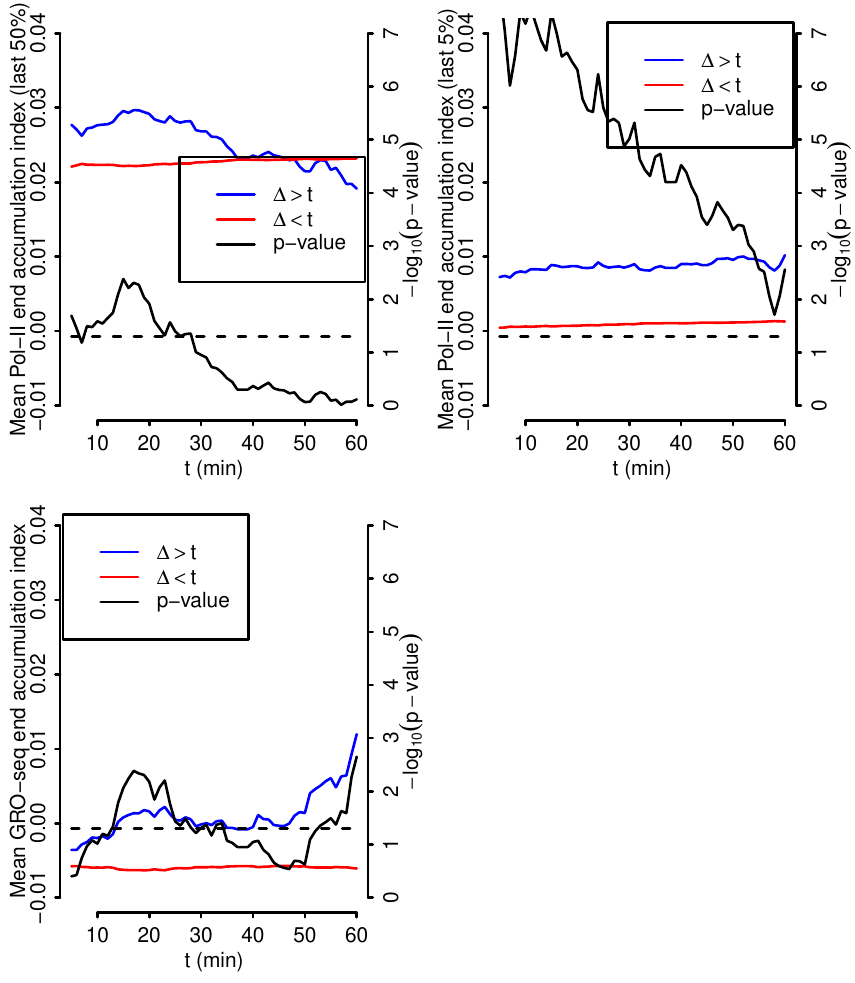}
  \caption{Alternative version of Fig.~\ref{fig:premrna_accumulation} (right) of the main paper:
    computing the index based on pol-II ChIP-seq and GRO-seq instead of intronic
    RNA-seq reads.  The top plots show differences in the mean pol-II
    accumulation index in long delay genes (blue) and short delay
    genes (red) as a function of the cut-off used to distinguish the
    two groups (left axis). Positive values indicate increased pol-II
    accumulation at the 3' end (top left: last 50\% of the gene body,
    top right: last 5\% of the gene body)
    over time. The black line shows the $p$-values of
    Wilcoxon's rank sum test between the two groups at each cut-off
    (right axis).
    The bottom plot is the same as top right, except for GRO-seq data
    of~\cite{Hah2011}, with the index is defined as the difference
    between the only late (160 min) time point and the average of the
    early (0-40 min) time points.
    In contrast to the pre-mRNA figure in the main paper, both long
    and short delay genes show a clear tendency towards accumulation
    of pol-II towards the end of the gene, but there is no clear
    difference between the two groups for the last 50\% (top left),
    while there is a very consistent pattern of more pol-II
    accumulation very close to 3' end (top right) for long delay genes,
    and the level is essentially independent of the estimated delay.
    GRO-seq data in the last 5\% (bottom) behave similarly as pol-II
    ChIP-seq (top right).
    \label{fig:pol2_accumulation}}
\end{figure}

In this section we provide supplementary
Figs.~\ref{fig:synthetic_halflives}--\ref{fig:pol2_accumulation}
discussed in the main paper as well as
Figs.~\ref{fig:rnastats}--\ref{fig:iqrboundplots} discussed in the
Supplementary Methods.


\subsection{Estimation of delays under changing mRNA half life}

\begin{table}[ht]
\centering
\caption{Fraction of reads mapping to mRNA transcripts alone (junction reads),
  pre-mRNA transcripts alone and both across all time points.
\label{tab:alignmentstats}}
\begin{tabular}{rrrr}
  \hline
  $t$ & mRNA & pre-mRNA & both \\ 
  \hline
  0 min & 0.035 & 0.287 & 0.678 \\ 
  5 min & 0.033 & 0.308 & 0.659 \\ 
  10 min & 0.036 & 0.270 & 0.694 \\ 
  20 min & 0.033 & 0.321 & 0.647 \\ 
  40 min & 0.033 & 0.312 & 0.654 \\ 
  80 min & 0.031 & 0.372 & 0.597 \\ 
  160 min & 0.033 & 0.331 & 0.636 \\ 
  320 min & 0.033 & 0.327 & 0.640 \\ 
  640 min & 0.034 & 0.330 & 0.636 \\ 
  1280 min & 0.032 & 0.339 & 0.629 \\ 
   \hline
\end{tabular}
\end{table}

\begin{table}[ht]
\centering
\caption{Fraction of reads assigned by BitSeq on average to mRNA
  transcripts and
  pre-mRNA transcripts, as well as fraction predicted for pre-mRNA
  when distributing 'both' category reads from
  Table~\ref{tab:alignmentstats} uniformly according to average
  transcript lengths. Only multi-exon genes are considered here,
  because the division is not meaningful for others. The results
  demonstrate that BitSeq can split the RNA-seq data to mRNA and
  pre-mRNA fractions in a meaningful manner.
\label{tab:bitseqstats}}
\begin{tabular}{rrrr}
  \hline
  $t$ & mRNA & pre-mRNA & pre-mRNA pred.\ \\ 
  \hline
  0 min & 0.62 & 0.38 & 0.37 \\ 
  5 min & 0.59 & 0.41 & 0.39 \\ 
  10 min & 0.64 & 0.36 & 0.36 \\ 
  20 min & 0.58 & 0.42 & 0.40 \\ 
  40 min & 0.59 & 0.41 & 0.40 \\ 
  80 min & 0.53 & 0.47 & 0.45 \\ 
  160 min & 0.57 & 0.43 & 0.41 \\ 
  320 min & 0.57 & 0.43 & 0.41 \\ 
  640 min & 0.57 & 0.43 & 0.41 \\ 
  1280 min & 0.55 & 0.45 & 0.42 \\ 
   \hline
\end{tabular}
\end{table}

\begin{figure}[tp]
  \centering
  \includegraphics{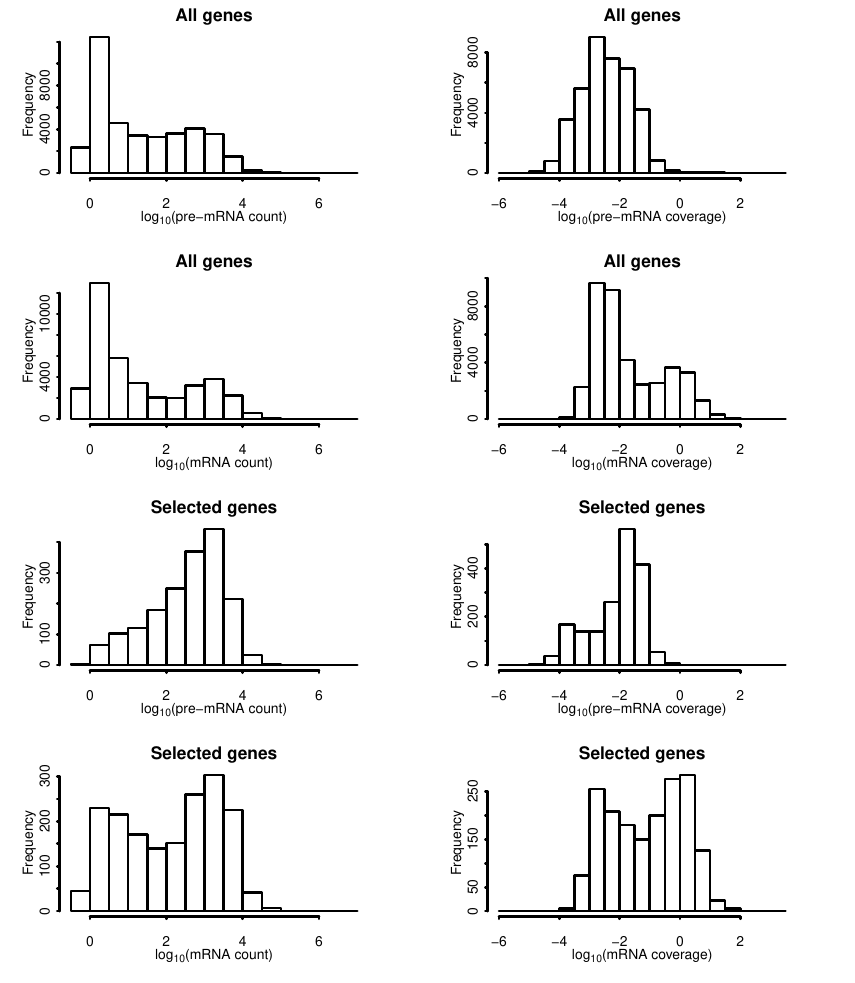}
  \caption{Distributions of per-gene mRNA and pre-mRNA counts and
    coverages based on BitSeq expression estimates. Top two rows show
    broad distributions for all genes, while the bottom two rows show
    distributions biased toward higher values for the selected 1786
    genes. The results show that the mRNA coverages are mostly
    clearly higher than for pre-mRNA, again demonstrating a sensible
    split between pre-mRNA and mRNA.
  \label{fig:rnastats}}
\end{figure}


\begin{figure}[tp]
  \centering
  \includegraphics{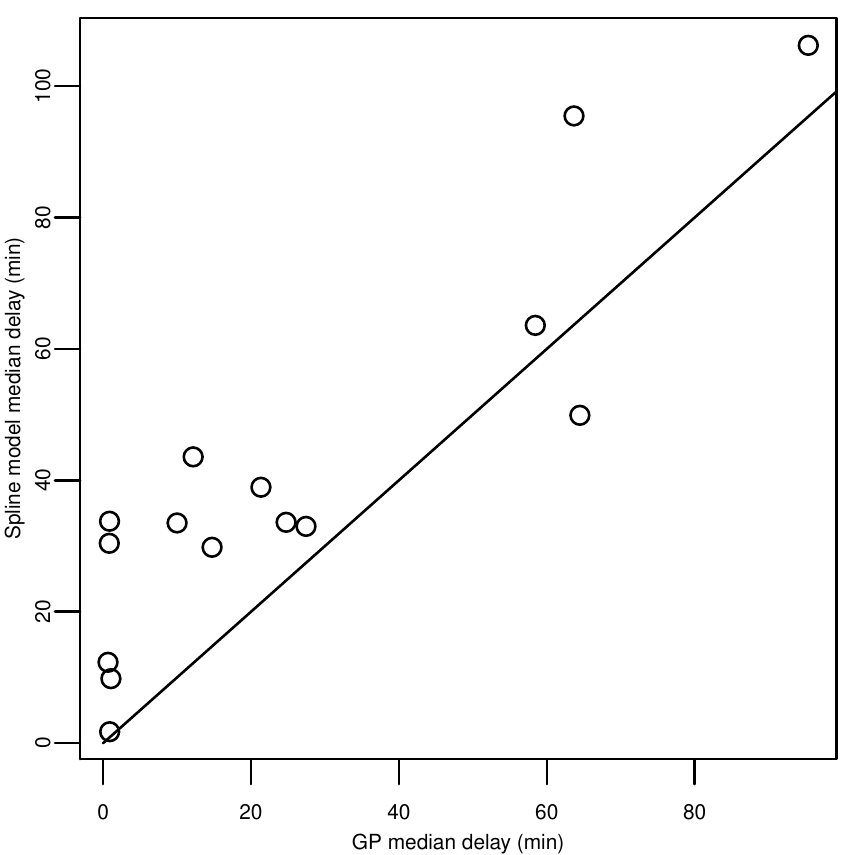}
  \caption{Comparison of estimated posterior median delays from the GP
    model with an alternative spline-based model for $n=15$ genes with
    reliable estimates. In this model we used cubic smoothing splines
    to fit continuous curves to pol-II measurements. To account for
    the uneven sampling the times were transformed as $t' =
    \log(t/\text{min}+5)$. The regularisation strength was shared over
    all genes and optimised by leave-one-out cross validation over all
    internal time points. The time transformation was also found to
    work much better than untransformed time in the cross
    validation. The smoothed pol-II curves were used as input to
    Eq.~\eqref{eq:differential_equation_appendix} which was solved numerically
    to obtain predictions for $m(t)$. Assigning a Gaussian noise model
    to $m(t)$ similar to the GP model and using similar priors for all
    shared parameters, we run MCMC to obtain posteriors over the
    parameters. We were only able to obtain reliable parameter
    estimates for a small subset of genes for which the method had a
    good fit (measured through expected relative residual variance)
    for both pol-II and mRNA. The other estimates were unreliable
    presumably because the method estimated the pol-II profiles
    independently but then ignored the uncertainty related to this
    estimation, which further highlights the benefits of the GP
    approach.
  \label{fig:splinecomparison}}
\end{figure}


\begin{figure*}[tp]
  \centering
  \includegraphics{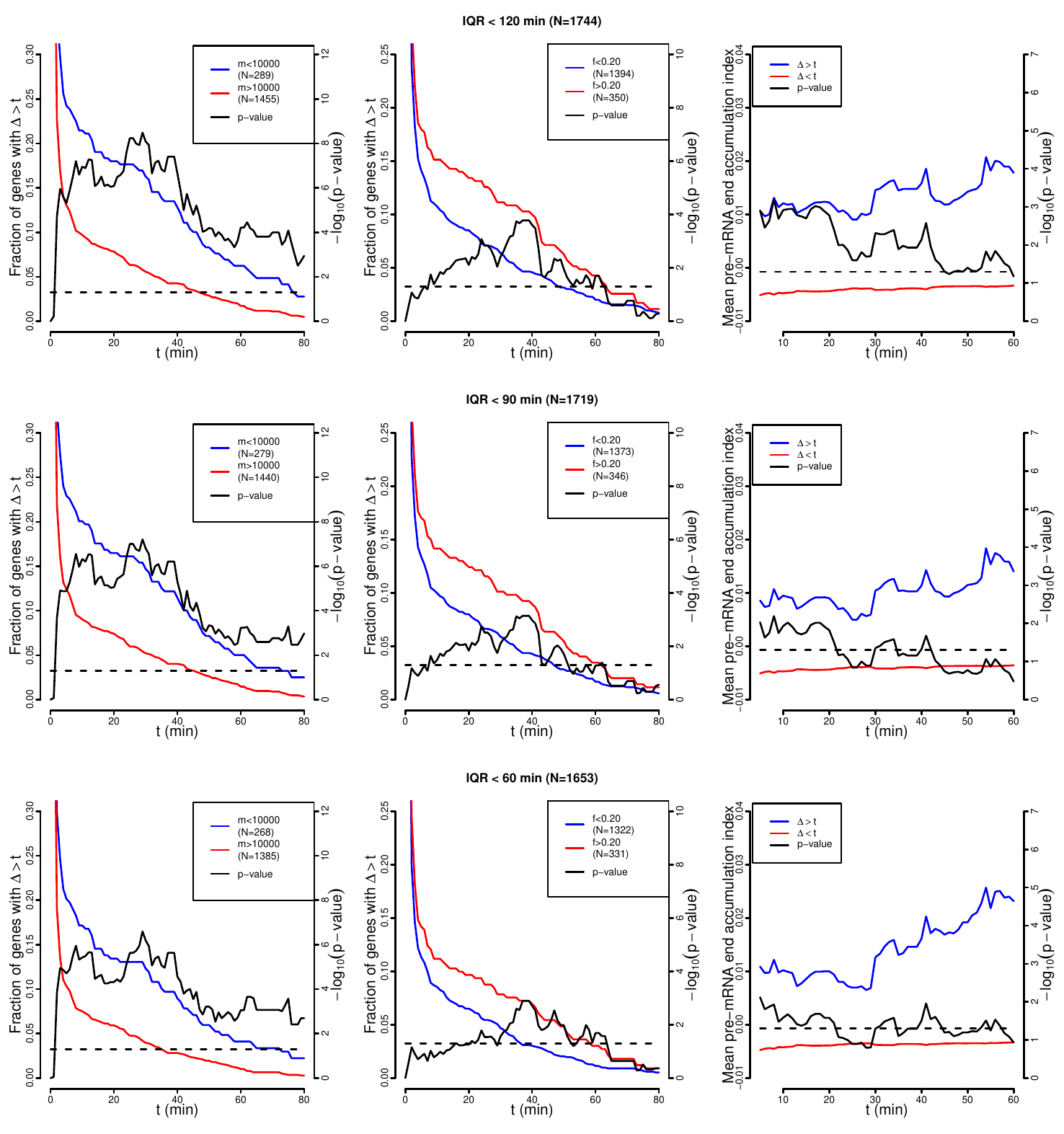}
  \caption{Alternative versions of Figs.~\ref{fig:delay_tails} and
    \ref{fig:premrna_accumulation} under more stringent filtering of
    delay posterior interquartile range (IQR).
  \label{fig:iqrboundplots}}
\end{figure*}

\bibliographystyle{myunsrt}
\bibliography{pol2rna_refs}

\end{document}